%% file: main.tex
\pdfoutput=1
\documentclass[12pt,a4paper]{article}
\usepackage{amsmath,amssymb}
\usepackage{graphicx}
\usepackage{booktabs,enumitem,microtype,placeins,float}
\usepackage{tikz}
\usetikzlibrary{arrows.meta,bending,decorations.markings,calc,positioning}
\usepackage{ifthen}
\usepackage{cite}
\usepackage[colorlinks=true,linkcolor=blue,citecolor=blue,urlcolor=blue]{hyperref}

\usepackage{newtxtext,newtxmath}
\DeclareMathAlphabet{\mathcal}{OMS}{cmsy}{m}{n}

\renewcommand{\thefootnote}{\fnsymbol{footnote}}
\newcounter{aff}
\renewcommand{\theaff}{\fnsymbol{aff}}
\newcommand{\affiliation}[1]{\setcounter{aff}{#1} $\rule{0em}{1.2ex}^\theaff\hspace{-.4em}$}

\setlist{itemsep=2pt,topsep=4pt,leftmargin=1.8em}
\allowdisplaybreaks[2]
\numberwithin{equation}{section}

\newcommand{\Kell}{\mathbf K}
\newcommand{\Eell}{\mathbf E}
\newcommand{\Piell}{\mathbf\Pi}
\hypersetup{pdftitle={Systematic High Overtone Expansion of Kerr Quasinormal Modes},pdfauthor={Yasuyuki Hatsuda}}

\begin{document}
\begin{titlepage}
\hfill\hfill
\begin{minipage}{1.2in}
RUP-26-30
\end{minipage}

\bigskip\bigskip\bigskip\bigskip
\begin{center}
{\LARGE\fontfamily{ptm}\fontseries{b}\selectfont Systematic High Overtone Expansion\\[2mm] of Kerr Quasinormal Modes}

\bigskip\bigskip
{\large Yasuyuki Hatsuda\footnote[1]{\tt yhatsuda@rikkyo.ac.jp}
}\\
\bigskip\bigskip
\affiliation{1}
{\normalsize\it Department of Physics, Rikkyo University, Toshima, Tokyo 171-8501, Japan}
\end{center}

\begin{abstract}
We develop a systematic method to compute the high overtone expansion of Kerr quasinormal modes on normal angular branches to arbitrary order. We first obtain the expansion of the angular eigenvalue at large frequencies by a simple polynomial recursion. Using this result in an alternative form of the radial Teukolsky equation, we find that all the WKB corrections are defined on a fixed genus-one spectral curve, and that the quantum periods at the second and higher orders are expressed in terms of only two elliptic periods. This allows us to determine the expansion coefficients order by order. We confirm that the truncated expansions reproduce quasinormal frequencies obtained by solving the angular and radial equations directly, for scalar, electromagnetic and gravitational fields. We also compare the scalar results with the exact WKB analysis of extremal Kerr black holes.
\end{abstract}

\end{titlepage}

\renewcommand{\thefootnote}{\arabic{footnote}}
\setcounter{footnote}{0}

\tableofcontents
\newpage
\section{Introduction}
\label{sec:introduction}
Quasinormal modes (QNMs) describe the damped oscillations of perturbed black holes, and they are a central subject in black hole perturbation theory~\cite{KokkotasSchmidt,BCS2009,KonoplyaZhidenko}. Their complex frequencies are determined by demanding that waves only fall into the horizon and only escape to spatial infinity. For Kerr black holes, the Teukolsky equation separates into angular and radial equations~\cite{Teukolsky}, and the QNM frequencies follow from the two coupled eigenvalue problems~\cite{Leaver}.

In this paper, we are interested in QNMs with high overtone numbers, i.e.\ with large negative imaginary parts of the frequency. For Schwarzschild black holes, Nollert computed such highly damped modes numerically and found that their real parts approach a constant~\cite{Nollert1993,Andersson1993}. Hod identified this constant as $T_H\log 3$, where $T_H$ is the Hawking temperature, and, by Bohr's correspondence principle, related it to the spacing of the quantized horizon area~\cite{Hod1998}, which was also discussed in connection with the black hole entropy~\cite{Dreyer2003}. This motivated analytic studies of the asymptotic spectrum by the monodromy method and complex WKB analysis~\cite{MotlSchwarzschild,MotlNeitzke,AnderssonHowls}, and they have been extended to non-asymptotically flat spacetimes and higher dimensions~\cite{CardosoNatarioSchiappa,NatarioSchiappa}. The first correction, of order $n^{-1/2}$ in four dimensions, was also obtained~\cite{Maassen,MusiriSiopsis,CardosoLemosYoshida}, and the analysis was extended to the next order~\cite{KaoSchwarzschild}.

More recently, exact WKB analysis has been applied to QNMs of Schwarzschild and extremal black holes~\cite{MNNO,HS}, and the high overtone spectrum was shown to be very sensitive to deformations of the potential~\cite{MNNOInstability}. Moreover, non-analytic subleading corrections in the large overtone expansion were shown to encode the geometry near a spacelike singularity inside the black hole~\cite{HartnollZhiboedov,XiaoNianLi}. These developments motivate us to go beyond the leading order.

For Kerr black holes, the problem is more involved, because the angular separation constant depends on the unknown complex frequency. Highly damped Kerr modes were studied numerically in~\cite{BCKO2003,BCY2004}, partly to test the extension of Hod's proposal to rotating black holes~\cite{Hod2003}. In particular, it was found in~\cite{BCY2004} that the leading real part at fixed $m$ is independent of $s$ and $\ell$, and corrections in inverse powers of the damping were suggested. The leading quantization conditions were then derived by complex WKB analysis~\cite{KH,KN}, and written in terms of elliptic integrals~\cite{KT}. The large-$c$ expansion of the angular eigenvalue, which is needed as an input, has also been studied~\cite{CO,BCC,VC}. It is known that the angular eigenvalue has normal branches, on which the separation constant grows linearly in $c$, and anomalous branches, on which its leading term can be quadratic in $c$~\cite{VC}. These works determine the leading behavior of the spectrum. However, it has not been known how to compute the higher-order corrections systematically. This paper aims to fill this gap.

We use $\chi=a/M$ and $c=a\omega$, with the complex frequency $\omega$, the mass $M$ and the specific angular momentum $a$, and consider $0<\chi<1$. For a fixed normal angular branch, we will show that the frequency has the expansion
\begin{equation}
2M\omega_N(\chi) \sim d_{-1}(\chi)\left(N+\frac12\right)+d_0(\chi)
       +\sum_{j\ge1}\frac{d_j(\chi)}{(N+\frac12)^j},\qquad N\to\infty,
 \label{eq:intro-series}
\end{equation}
and that all the coefficients $d_j(\chi)$ can be determined analytically.\footnote{A Mathematica package that computes the coefficients $d_j$ up to $j=24$ and evaluates~\eqref{eq:intro-series} for the normal branches is available at Zenodo, \href{https://doi.org/10.5281/zenodo.23691045}{doi:10.5281/zenodo.23691045}.} Here $N$ is an integer. Since the coefficients $d_j(\chi)$ are determined independently of the modes,~\eqref{eq:intro-series} itself fixes the integer $N$ for each highly damped mode. As explained below, $N$ is closely related to the overtone number $n$. Note that we consider the limit of large damping at fixed $\ell$, not the eikonal limit of large $\ell$.

There is another approach to the QNM problem. The connection between black hole QNMs and four-dimensional $\mathcal N=2$ gauge theories was first found in~\cite{AGH}. In particular, for Kerr black holes, the two separated Teukolsky equations can be identified with quantized Seiberg--Witten curves of $SU(2)$ gauge theory with three flavors, and the QNM frequencies obey exact quantization conditions written in terms of the free energy in the Nekrasov--Shatashvili limit~\cite{NS2009}. See also~\cite{BILT2022} for a related approach based on two-dimensional CFT. In practice, this method gives accurate frequencies of low overtones, although the instanton series converges rather slowly~\cite{AGH}. Moreover, it was observed in~\cite{ArnaudoGrassiHao} that, for some black hole problems, the instanton counting parameters can be of order one, and the corresponding point of the Seiberg--Witten moduli space is at strong coupling. An expansion around a strong coupling point, rather than the instanton expansion, might improve the convergence at high overtones, but to our knowledge such an expansion is not available for the Kerr QNM problem. In this paper, we instead study the high overtone regime directly from the differential equations.

Our idea is as follows. At fixed rotation $0<\chi<1$, the parameter $c$ becomes large at high overtones. We first compute the angular eigenvalue for large $c$. On the normal branch, the problem reduces to a harmonic oscillator near the equator, and a simple polynomial recursion, similar to the Bender--Wu method~\cite{BW}, gives its expansion to arbitrary order. We then insert this expansion into the dual radial equation proposed by the author~\cite{HatsudaAlt}, which looks much simpler than the radial Teukolsky equation. It turns out that the leading WKB curve is a fixed genus-one curve, which depends only on the spin $\chi$ and not on $s$, $m$ or $\ell$. Higher-order WKB corrections change the differential, but not the curve. Moreover, the quantum periods at the second and higher orders reduce to only two elliptic periods. A similar structure appears in the extremal case~\cite{HS}, where every quantum correction to the period is a combination of the complete elliptic integrals $\Kell$ and $\Eell$. Finally, the inversion of the quantization condition gives the coefficients $d_j(\chi)$ order by order, without any numerical fitting.

We test our expansion in two ways. First, we compare it with QNM frequencies obtained by solving the angular and radial equations directly, for the scalar, electromagnetic and gravitational cases. We find that, at $\chi=4/5$ and $N=120$, the truncated expansion reproduces the QNM frequencies with errors of order $10^{-21}$ without resummation. Second, we compare the scalar results with the exact quantization condition for extremal Kerr black holes in~\cite{HS}. At the classical level, the two periods are related analytically, and the frequency coefficients also agree beyond the leading order in the tested cases. We also find that the coefficients grow factorially, and that the inverse overtone corrections become singular in the small spin limit.

Let us summarize the assumptions in this paper. We consider only the normal angular branches. We do not consider the anomalous branches~\cite{VC}, or the highly damped sector found numerically in~\cite{KBM}, whose angular behavior is similar to the oblate case. For $m=0$, the expansion~\eqref{eq:intro-series} is purely imaginary to all orders, and its relation to the physical modes requires care at the negative imaginary axis, where the boundary condition is defined by a one-sided limit. We do not examine the axisymmetric sequences in this paper, and our direct tests concern $m=-2$. We use the dual radial equation~\cite{HatsudaAlt} instead of the radial Teukolsky equation. Their isospectrality in the whole lower half plane has not been proven rigorously, but it is strongly supported analytically and numerically (see Section~\ref{sec:dual-radial} and Appendix~\ref{app:dual}), and we assume it in this paper. We also assume that the all-order extension of the quantization condition in~\cite{KH,KN} holds, and we do not discuss possible nonperturbative corrections. The numerical tests in Section~\ref{sec:normal-checks} give strong support to these assumptions. The standard and dual radial equations give the same QNM frequencies, and the error of the truncated expansion behaves as expected for an asymptotic series. Finally, the integer $N$ in~\eqref{eq:intro-series} labels the quantization condition~\eqref{eq:normal-quantization}. It is not the overtone number $n$ in general, and their difference depends on the mode (see Appendix~\ref{app:normal-numerics}).

The rest of this paper is organized as follows. Section~\ref{sec:angular} gives the large-$c$ expansion of the angular eigenvalue. In Section~\ref{sec:normal-radial}, we construct the all-order WKB expansion of the radial problem and determine the frequency coefficients. We also discuss their rotation dependence. In Section~\ref{sec:normal-checks}, we test our results against numerical QNM frequencies and the exact WKB analysis of extremal Kerr black holes. We conclude in Section~\ref{sec:conclusion}. Technical details are collected in Appendices~\ref{app:analytic} and \ref{app:numerical}. The Borel structure of the series will be discussed elsewhere.
\section{Angular eigenvalue problem and normal prolate expansion}
\label{sec:angular}
We first compute the angular eigenvalue as a function of $c$, before imposing the radial quantization condition. We fix $s$ (the spin weight) and $m$ (the azimuthal number). Below we consider $s=0,-1,-2$, that is, scalar, electromagnetic and gravitational fields. The angular wave function $S(x)$ and separation constant $A(c)$ obey the standard angular Teukolsky equation~\cite{Teukolsky}
\begin{equation}
 \frac{d}{dx}\left[(1-x^2)\frac{dS}{dx}\right]
 +\left[c^2x^2-2csx+A(c)+s-\frac{(m+sx)^2}{1-x^2}\right]S=0,
 \qquad x=\cos\theta.
 \label{eq:angular-standard}
\end{equation}
Here $\theta$ is the polar angle. The physical endpoint condition selects the regular Frobenius solution at each of $x=\pm1$. If $k_\pm=|m\pm s|/2$, these solutions behave as $(1-x)^{k_+}$ and $(1+x)^{k_-}$, respectively. When $k_+=0$ or $k_-=0$, the two exponents at the corresponding endpoint coincide, and the second solution, which contains $\log(1\mp x)$, is excluded. At $c=0$, $A=\ell(\ell+1)-s(s+1)$ defines the spherical harmonic label $\ell$. At complex $c$, we transport this label continuously from $c=0$.

It is useful to state the exact spectral condition before making an approximation. Let $S^{\rm reg}_-$ and $S^{\rm reg}_+$ be non-zero regular solutions selected at the left and right endpoints and continued to the same interior point. They represent one solution regular at both ends precisely when
\begin{equation}
 (1-x^2)W[S^{\rm reg}_-,S^{\rm reg}_+]=0,
 \qquad W[f,g]=fg'-f'g.
 \label{eq:angular-wronskian}
\end{equation}
Here the prime denotes the derivative with respect to $x$. By Abel's identity, $(1-x^2)W[S^{\rm reg}_-,S^{\rm reg}_+]$ is independent of $x$. All angular approximations below describe zeros of this condition.

\subsection{Equatorial scaling}
The prolate regime is selected naturally by the high overtone limit considered here. At fixed non-extremal rotation $0<\chi=a/M<1$, the normal high damping sequence has $\operatorname{Im}\omega\to-\infty$, and hence $c=a\omega=\chi\Omega/2$, with $\Omega=2M\omega$, becomes large with its leading asymptotic direction approaching the negative imaginary axis. We therefore first construct the angular expansion on the pure prolate ray $c=-i C$, $C>0$, $C\to+\infty$, and subsequently continue it to the nearby complex $c$ trajectory relevant to the QNM branch. Note that this argument fixes only the direction of $c$. The angular branch has to be chosen separately. Throughout this paper, we restrict ourselves to normal branches. In the numerical tests, the branch is transported continuously from the Schwarzschild limit while $\chi$ is varied, so that $c=\chi\Omega/2$ follows the induced complex path.

For the scalar spheroidal equation, an all-order WKB framework based on quantum actions and Picard--Fuchs equations, including perturbative and resurgent spectral information, was recently developed in~\cite{MeynigSpheroidal}. Our construction instead starts directly from the spin-weighted angular Teukolsky equation, and gives a simple polynomial recursion for the large-$c$ expansion.

Normal prolate sequences are obtained by setting $c=-i C$, with $C>0$ large, and looking for $A=O(C)$. The equatorial width is $x=O(C^{-1/2})$, since the leading angular terms balance as $-C^2x^2+A$. We remove the first derivative by setting $S=\psi/\sqrt{1-x^2}$. The normal form is
\begin{equation}
 \psi''+U(x;c,A)\psi=0,\qquad
 U=\frac{c^2x^2-2csx+A+s}{1-x^2}
       +\frac{1-(m+sx)^2}{(1-x^2)^2}.
 \label{eq:angular-normal-form}
\end{equation}
We denote the corresponding normal form solutions by $\psi_\pm^{\rm reg}=\sqrt{1-x^2}\,S_\pm^{\rm reg}$. In particular $W[\psi_-^{\rm reg},\psi_+^{\rm reg}]=(1-x^2)W[S_-^{\rm reg},S_+^{\rm reg}]$ does not depend on $x$.

We put $\varepsilon=C^{-1/2}$ and $x=\varepsilon y$, and look for
\begin{equation}
 A=\eta\varepsilon^{-2}+\sum_{j\ge0}\mathcal A_j^{\rm nor}\varepsilon^{2j}.
 \label{eq:normal-angular-series}
\end{equation}
Here $\eta$ and the coefficients $\mathcal A_j^{\rm nor}$ are initially unknown and are fixed order by order by the solvability conditions below. Multiplication of the equation by $\varepsilon^2$ gives
\begin{equation}
 \left[\partial_y^2+\eta-y^2+\sum_{r\ge1}\varepsilon^r V_r(y)\right]\psi=0.
 \label{eq:angular-scaled}
\end{equation}
Every $V_r$ is determined by expanding the following rational expression at $\varepsilon=0$:
\begin{equation}
 \eta-y^2+\sum_{r\ge1}\varepsilon^rV_r
 =\frac{\eta-y^2+2i s\varepsilon y+s\varepsilon^2+
             \sum_{j\ge0}\mathcal A_j^{\rm nor}\varepsilon^{2j+2}}
            {1-\varepsilon^2 y^2}
 +\frac{\varepsilon^2[1-(m+s\varepsilon y)^2]}{(1-\varepsilon^2y^2)^2}.
 \label{eq:Vr-generator}
\end{equation}
Therefore no additional input is needed at higher orders. For example,
\begin{align}
 V_1&=2i sy,\qquad
 V_2=\mathcal A_0^{\rm nor}+\eta y^2-m^2+s+1-y^4,\nonumber\\
 V_3&=2sy(-m+i y^2).
 \label{eq:Vr-low}
\end{align}
The leading solution that decreases on both sides of the equatorial well is
$e^{-y^2/2}H_{\bar L}(y)$, with $\eta=2\bar L+1$ and integer $\bar L\ge0$. Here $H_p$ is the Hermite polynomial of degree $p$. This local oscillator level labels a normal asymptotic sequence. For $s=0$, $\bar L=\ell-|m|$. For $s\ne0$, the connection to a spherical label must be established separately. According to the numerical classification in~\cite{VC} along the negative imaginary $c$ axis, one has $\bar L=\ell-\max(|m|,|s|)$ for $|s|\le2$, except for $|s|=|m|=2$, where the sequence with $\ell=3$ is anomalous and $\bar L=\ell-3$ for $\ell\ge4$. Table~\ref{tab:Lbar} lists the values for small $\ell$.
\begin{table}[tbp]
 \centering
 \caption{The label $\bar L$ of the normal sequences for $s=-1,-2$ and small $\ell$, obtained from the classification in~\cite{VC}. It depends only on $|m|$. A dash means that the mode does not exist ($\ell<\max(|m|,|s|)$), while ``a'' means that the mode exists but its sequence is anomalous, so that $\bar L$ is not defined. For $s=0$, $\bar L=\ell-|m|$.}
 \label{tab:Lbar}
 \begin{tabular}{c cccccc c cccccc}
 \toprule
 & \multicolumn{6}{c}{$s=-1$} & & \multicolumn{6}{c}{$s=-2$}\\
 \cmidrule{2-7}\cmidrule{9-14}
 $\ell$ & $|m|=0$ & 1 & 2 & 3 & 4 & 5 & & $|m|=0$ & 1 & 2 & 3 & 4 & 5\\
 \midrule
 1 & 0 & 0 & -- & -- & -- & -- & & -- & -- & -- & -- & -- & --\\
 2 & 1 & 1 & 0 & -- & -- & -- & & 0 & 0 & 0 & -- & -- & --\\
 3 & 2 & 2 & 1 & 0 & -- & -- & & 1 & 1 & a & 0 & -- & --\\
 4 & 3 & 3 & 2 & 1 & 0 & -- & & 2 & 2 & 1 & 1 & 0 & --\\
 5 & 4 & 4 & 3 & 2 & 1 & 0 & & 3 & 3 & 2 & 2 & 1 & 0\\
 \bottomrule
 \end{tabular}
\end{table}

\subsection{Polynomial recursion and the solvability condition}
We factor out the Gaussian and expand
\begin{equation}
 \psi=e^{-y^2/2}\sum_{r\ge0}\varepsilon^r P_r(y),\qquad
 P_0=H_{\bar L},\qquad
 \mathcal D_{\bar L}=\partial_y^2-2y\partial_y+2\bar L.
 \label{eq:angular-P}
\end{equation}
At order $r$, the equation is
\begin{equation}
 \mathcal D_{\bar L}P_r=-\sum_{k=1}^r V_kP_{r-k},\qquad
 \mathcal D_{\bar L}H_p=2(\bar L-p)H_p.
 \label{eq:hermite-recursion}
\end{equation}
This polynomial recursion is the analogue of the Bender--Wu construction for the anharmonic oscillator~\cite{BW}. We expand the right-hand side in Hermite polynomials. Since $\mathcal D_{\bar L}$ annihilates $H_{\bar L}$, the coefficient of $H_{\bar L}$ on the right-hand side must vanish. At order $r=2j+2$, this condition determines the eigenvalue coefficient $\mathcal A_j^{\rm nor}$, which first appears at this order. For $p\neq\bar L$, the coefficient of $H_p$ in $P_r$ is obtained by dividing that of the right-hand side by $2(\bar L-p)$. The coefficient of $H_{\bar L}$ in $P_r$ is not fixed by the equation, and we set it to zero for $r\ge1$, as in the intermediate normalization of the Rayleigh--Schr\"odinger perturbation theory. This choice only fixes the normalization of $\psi$, and does not affect $A$. Parity implies $P_r(-y)=(-1)^{\bar L+r}P_r(y)$, so the condition is automatically satisfied at odd orders.

Let us see how the first two steps work. At first order, the recursion gives $\mathcal D_{\bar L}P_1=-V_1P_0=-2i s y H_{\bar L}$.
The identity $yH_p=H_{p+1}/2+pH_{p-1}$ gives the first wave function correction,
\begin{equation}
 P_1=\frac{i s}{2}H_{\bar L+1}-i s\bar L H_{\bar L-1}.
 \label{eq:normal-P1}
\end{equation}
Let $[F]_{\bar L}$ denote the coefficient of $H_{\bar L}$ in the Hermite expansion of a polynomial $F$. At order $\varepsilon^2$,
\begin{equation}
 \mathcal D_{\bar L}P_2=-\left(V_1P_1+V_2H_{\bar L}\right),
 \label{eq:normal-P2}
\end{equation}
and the right-hand side must have vanishing $H_{\bar L}$ component. Writing $q=\bar L(\bar L+1)$, the required diagonal projections are
\begin{align}
 [y^2H_{\bar L}]_{\bar L}=\bar L+\frac12,\qquad[y^4H_{\bar L}]_{\bar L}=\frac34(2q+1),\qquad [V_1P_1]_{\bar L}=-s^2.
 \label{eq:normal-A0-projections}
\end{align}
The solvability condition therefore becomes
\begin{equation}
 0=[V_1P_1+V_2H_{\bar L}]_{\bar L}
   =\mathcal A_0^{\rm nor}+\frac q2-m^2+s+\frac34-s^2,
 \label{eq:normal-A0-solvability}
\end{equation}
which determines $\mathcal A_0^{\rm nor}$.
This calculation also shows explicitly the different roles of the two conditions in the recursion: the resonant projection determines the eigenvalue coefficient, whereas the homogeneous $H_{\bar L}$ component of $P_r$ is fixed only by the normalization convention. Continuing the same operations gives
\begin{align}
 \mathcal A_0^{\rm nor}&=m^2-\frac q2+s^2-s-\frac34,\nonumber\\
 \mathcal A_1^{\rm nor}&=\frac{\eta}{16}(8m^2-q-3+16s^2)+2i ms^2,\nonumber\\
 \mathcal A_2^{\rm nor}&=-\frac{5q^2+35q+15}{64}
 +\left(\frac{3q}{2}-2s^2+\frac34\right)m^2
 +4i\eta ms^2+3qs^2-s^4+\frac32s^2.
 \label{eq:normal-angular-low}
\end{align}
Equations~\eqref{eq:Vr-generator} and~\eqref{eq:hermite-recursion}, including their projection and normalization, define the arbitrary finite order algorithm. The coefficients up to $\mathcal A^{\rm nor}_{39}$ are included in the Mathematica package described in Data availability. Our results agree with the fits in~\cite{VC} at low orders.

For the gravitational family $(s,\bar L)=(-2,0)$, retaining the signed $m$ makes the first terms particularly simple:
\begin{align}
 A(-i C)={}&C+m^2+\frac{21}{4}
 +\frac{(8m^2+61)/16+8i m}{C}\nonumber\\
 &+\frac{-29m^2/4+16i m-655/64}{C^2}+O(C^{-3}).
 \label{eq:gravity-angular-example}
\end{align}
The normal gravitational tests below use $m=-2$. For the electromagnetic family $(s,\bar L)=(-1,0)$, retaining the signed $m$ gives
\begin{equation}
 A(-i C)=C+m^2+\frac54
 +\frac{(8m^2+13)/16+2i m}{C}
 +\frac{17/64-(5/4)m^2+4i m}{C^2}+O(C^{-3}).
 \label{eq:EM-angular-example}
\end{equation}
The electromagnetic numerical benchmark below uses $m=-2$, which reproduces the specialized coefficients $21/4$, $45/16-4i$ and $-303/64-8i$. For a scalar field, $\bar L=\ell-|m|$ and the expansion reads
\begin{align}
 A(-i C)&=\eta C+a_0+\frac{b_1}{C}+\frac{b_2}{C^2}+\cdots,\nonumber\\
 a_0&=m^2-\frac{\eta^2+5}{8},\qquad
 b_1=-\frac{\eta(\eta^2+11-32m^2)}{64},\nonumber\\
 b_2&=-\frac{5(\bar L^4+2\bar L^3+8\bar L^2+7\bar L+3)-48m^2(2\bar L^2+2\bar L+1)}{64}.
 \label{eq:scalar-b2}
\end{align}
This includes the scalar coefficient needed in the extremal comparison for $\bar L>0$.\footnote{In the notation of~\cite{MeynigSpheroidal}, $\hbar_{\rm M}=i\sqrt2/c$ and $\mu_{\rm M}=(1/4-m^2)/c^2$, so that $\mu_{\rm M}=\tfrac12(m^2-1/4)\hbar_{\rm M}^2\to0$ at fixed $m$ and large $c$. After matching the conventions and using the symmetry $c\mapsto-c$, the perturbative results in~\cite{MeynigSpheroidal} reproduce~\eqref{eq:scalar-b2}, including $b_2$. The nonperturbative analysis in~\cite{MeynigSpheroidal} is done at fixed $\mu_{\rm M}$, and cannot be applied directly to our case.}

These formulas describe the normal balance $A=O(C)$. Note that~\eqref{eq:angular-wronskian} has other solutions as well. For non-zero spin, anomalous sequences can instead have a quadratic leading balance~\cite{VC}. In the following, we restrict ourselves to a normal branch. The numerical checks of the angular problem, including branch tracking and cutoffs, are described in Appendix~\ref{app:direct}.
\section{All-order WKB construction of the normal frequency expansion}
\label{sec:normal-radial}
In this section, we fix a normal angular branch and the radial integration cycle in the quantization condition. Since the leading WKB curve is nondegenerate for $0<\chi<1$, this cycle is continued uniquely along $\chi$. Our purpose is to turn the angular recursion into the frequency coefficients in~\eqref{eq:intro-series}. We first identify the leading curve and a convenient Riccati equation. We then explain why all higher periods reduce to the same two elliptic integrals. Finally, we invert the quantization condition, and examine the rotation dependence of the result.

\subsection{Dual radial equation and boundary channels}
\label{sec:dual-radial}
The dual radial equation proposed in~\cite{HatsudaAlt} reads
\begin{equation}
 \left[f(z)\frac{d}{dz}f(z)\frac{d}{dz}+\Omega^2-V_{\rm d}(z)\right]\phi(z)=0,
 \qquad f(z)=1-\frac1z,
 \label{eq:alt}
\end{equation}
with
\begin{equation}
 V_{\rm d}=f\left[4c^2+\frac{4c(m-c)}z
  +\frac{A(c)+s(s+1)-c(2m-c)}{z^2}-\frac{s^2-1}{z^3}\right].
 \label{eq:V}
\end{equation}
We refer to $z=1$ and $z=\infty$ as the horizon and infinity, because the boundary channels below are the images of the corresponding Teukolsky scattering channels. Note that $z$ is not a simple reparametrization of the Boyer--Lindquist radius. In the Schwarzschild limit, it was observed in~\cite{AGH} that the exchange of the Seiberg--Witten mass parameters $\mu_2\leftrightarrow\mu_3$ maps the Teukolsky dictionary to the Regge--Wheeler dictionary. Appendix~\ref{app:dual} explains the Kerr case using the symmetric confluent Heun equation, the same mass exchange and Whiting's integral transformation.\footnote{In the Kerr--de Sitter case, such mass symmetries were recently explained by a symmetry of the Painlev\'e VI equation through isomonodromic deformations, and the dual Lorentzian spacetime, on which a subset of them gives a wave equation, was found to represent a black hole again~\cite{HIZ}.}

The characteristic recurrence is invariant under the same permutation. Moreover, the mass swapped equation is the one obtained by Whiting's transformation in its mode stability domain $\operatorname{Im}\omega\ge0$~\cite{Whiting,TdC,CasalsTdC}. For QNM frequencies in the lower half plane, the isospectrality is strongly supported by the continuation of the characteristic recurrence and by direct numerical comparisons of the two radial formulations (see Appendix~\ref{app:normal-numerics}). As stated in Section~\ref{sec:introduction}, we assume it in this paper.

In this frame the horizon ingoing and infinity outgoing solutions have the asymptotic forms
\begin{align}
 \phi_H^{\rm in}&\sim(z-1)^{-i\Omega},&&z\to1,\label{eq:hor}\\
 \phi_\infty^{\rm out}&\sim e^{i\rho\Omega z}z^{i(\Omega-\chi m)/\rho}
               [1+O(z^{-1})],&&z\to\infty.
 \label{eq:inf}
\end{align}
Here $\rho=\sqrt{1-\chi^2}$. The power at infinity is a Coulomb-type factor, which is a part of the channel. We denote the complementary incoming channel at infinity by $\phi_\infty^{\rm in}$. Let
\begin{equation}
 \phi_H^{\rm in}=A_{\rm out}(\Omega)\phi_\infty^{\rm out}
                    +A_{\rm in}(\Omega)\phi_\infty^{\rm in}.
 \label{eq:radial-connection}
\end{equation}
The resonance condition is $A_{\rm in}=0$, equivalently the vanishing of the Wronskian between $\phi_H^{\rm in}$ and $\phi_\infty^{\rm out}$ at an interior point. The conserved radial Wronskian is $fW$ in the $z$-variable. These boundary solutions are defined by analytic continuation in $\Omega$, and their branch on the negative imaginary axis is fixed as in the continued fraction of Appendix~\ref{app:direct}. Once these two channels are fixed, we regard~\eqref{eq:alt} as a one-dimensional scattering problem, and apply the complex WKB formulation in~\cite{KH,KN}. In this formulation, QNM poles are associated perturbatively with the external turning point connection. We call the corresponding cycle the outer cycle below.\footnote{Here ``outer'' refers to the external scattering class in~\cite{KH,KN}, not to a region of the $z$-plane.}

\subsection{High frequency scaling and the fixed elliptic curve}
\label{sec:radial-geometry}
For the WKB analysis, we use $u=1/z$, and define
\begin{equation}
 \sigma=\chi^2,\qquad \hbar=\Omega^{-1},\qquad
 F(u)=u^2(u-1),\qquad D=F(u)\partial_u.
 \label{eq:radial-definitions}
\end{equation}
Here and below, the prime on a function of $u$ denotes the derivative with respect to $u$. Then $f\partial_z=D$, and~\eqref{eq:alt} becomes
\begin{equation}
 F^2\phi''+FF'\phi'+\Omega^2 Q(u;\hbar)\phi=0,
 \qquad Q=1-\hbar^2 V_{\rm d}(1/u).
 \label{eq:poly-ODE}
\end{equation}
To write~\eqref{eq:poly-ODE} in the Schr\"odinger form, we set $\Psi=\sqrt F\,\phi$ on a continued branch of the square root, and obtain
\begin{align}
& \hbar^2 \Psi''+\widehat Q\Psi=0,\nonumber\\
& \widehat Q=\frac Q{F^2}
 +\hbar^2\left[-\frac{F''}{2F}+\frac{(F')^2}{4F^2}\right]
 =\frac Q{F^2}+\hbar^2\frac{4-3u}{4u(u-1)^2}.
 \label{eq:u-normal}
\end{align}
Therefore we can use the ordinary normal form, while keeping the polynomial form~\eqref{eq:poly-ODE} for the calculation.

Since $\hbar=\Omega^{-1}$, we expand the polynomial potential as $Q(u;\hbar)=\sum_{j\ge0}\hbar^jQ_j(u)$. For the normal angular input, $A=O(\Omega)$, while $c^2=\sigma\Omega^2/4$. The terms of order $\Omega^2$ in $V_{\rm d}$ are multiplied by $\hbar^2$ in $Q=1-\hbar^2V_{\rm d}$ and therefore determine the leading coefficient of $Q$,
\begin{equation}
 Q_0(u)=\frac14G_\sigma(u),\qquad
 G_\sigma(u)=\sigma u^3-5\sigma u^2+8\sigma u+4(1-\sigma).
 \label{eq:G}
\end{equation}
Therefore the leading curve and its meromorphic differential are
\begin{equation}
 Y^2=G_\sigma(u),\qquad
 \lambda_0=\frac{Y}{2u^2(u-1)}\,du.
 \label{eq:leading-curve}
\end{equation}
The cubic discriminant is $-16\sigma^2(\sigma+27)$, which is non-zero for $0<\sigma\le1$. The compact curve has genus one. Its branch points are the three finite zeros of $G_\sigma$ and $u=\infty$. The lifts of $u=0,1$ are punctures of $\lambda_0$ for generic non-extremal spin. Interestingly, $G_\sigma$ is determined by $\sigma=\chi^2$ alone, and is independent of $s$, $m$ and $\bar L$. These mode dependent quantities enter only through the subleading terms $Q_j$ and hence through the higher WKB differentials on the same spectral curve. Therefore increasing the WKB order changes the differential, but not the underlying genus-one curve.

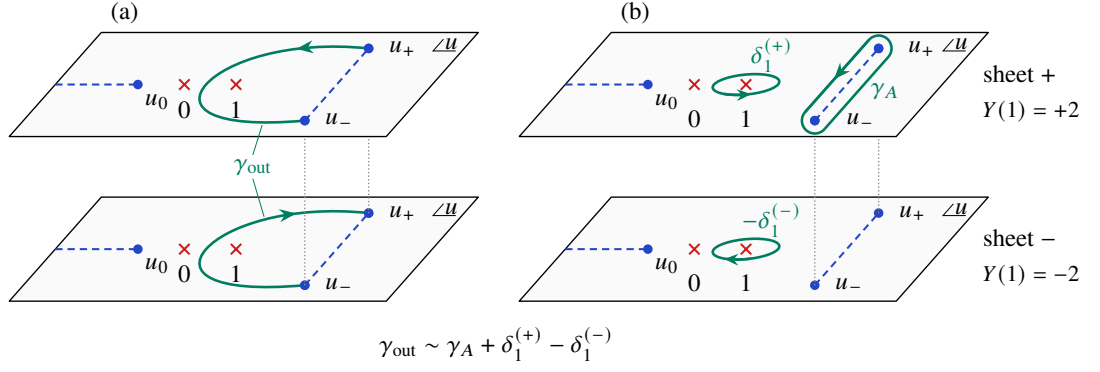
\begin{figure}[!tbp]
 \centering\resizebox{\textwidth}{!}{\input{two_sheet_cycle_twopanel.tikz}}
 \caption{The QNM cycle $\gamma_{\rm out}$ on the leading curve~\eqref{eq:leading-curve} at $\chi=1/2$. The two sheets are labeled by the value $Y(1)=\pm2$ at the lift of $u=1$. Dots, crosses and dashed lines denote the turning points, the punctures $u=0,1$ and the branch cuts, respectively, and dotted lines indicate how the sheets are glued along the cut. (a)~The cycle $\gamma_{\rm out}$ goes from $u_+$ to $u_-$ on sheet $+$ and returns on sheet $-$. (b)~The decomposition~\eqref{eq:outer-cycle}.}
 \label{fig:normal-curve}
\end{figure}

For numerical evaluation the roots can be written without solving a cubic at every order. With
$\tau_\chi=\frac23\operatorname{arsinh}(3\sqrt3/\chi)$, they are
\begin{equation}
 u_0=\frac53-\frac23\cosh\tau_\chi,\qquad
 u_\pm=\frac53+\frac13\cosh\tau_\chi
                  \pm\frac{i}{\sqrt3}\sinh\tau_\chi.
 \label{eq:turning-points}
\end{equation}
The external QNM turning point connection can be lifted to a closed cycle on the two-sheeted curve~\eqref{eq:leading-curve}: one follows the connection on one sheet and returns on the other. Let $\gamma_A$ denote the compact cycle surrounding the cut between $u_-$ and $u_+$, with orientation chosen so that the integral of $\lambda_0$ along $\gamma_A$ is negative imaginary. Because $\lambda_0$ has simple poles with non-zero residues at the two lifts of $u=1$, the QNM/external class on the punctured curve is not represented by $\gamma_A$ alone. For the branch cut convention in Figure~\ref{fig:normal-curve},
\begin{equation}
 \gamma_{\rm out}\sim \gamma_A+\delta_1^{(+)}-\delta_1^{(-)},
 \label{eq:outer-cycle}
\end{equation}
where $\delta_1^{(\pm)}$ are positively oriented small loops around the two lifts $p_1^{(\pm)}=(u=1,Y=\pm2)$. The right-hand side is a convenient homology decomposition. Figure~\ref{fig:normal-curve}(a) shows a connected representative on the glued two-sheeted Riemann surface, obtained by crossing the branch cut between the sheets, and Figure~\ref{fig:normal-curve}(b) shows the pieces of~\eqref{eq:outer-cycle}. Semiclassically, $\gamma_A$ carries the turning point action, while the puncture contribution supplies the horizon monodromy required by the QNM scattering class. This structure follows the analysis in~\cite{KH}, where the QNM condition is derived from a closed contour through two complex turning points, which encloses the event horizon as its only singular point. No additional loop about the lifts of $u=0$ is included in this class. The local monodromy at $u=0$, which represents spatial infinity, is already fixed by the Coulomb power in the selected outgoing channel~\eqref{eq:inf}. This is consistent with~\cite{KH,KN}, where the boundary condition at infinity is imposed by analytic continuation along an anti-Stokes line. Adding a further $u=0$ winding would compose the continuation with an extra infinity monodromy and hence define a different scattering/homology class. Similarly, different contours correspond to different families of resonances, such as total transmission and total reflection modes, in~\cite{KN}. Figure~\ref{fig:normal-curve} displays this convention schematically. We stress that the cycle is a part of the input, and it is not determined by the curve alone.
The compact curve stays nonsingular at $\chi=1$, where $u_0=0$ and $u_\pm=(5\pm i\sqrt7)/2$. A puncture meets a branch point there, however, and the extremal problem ($a=M$) must still be treated separately. At $\chi=0$, the compact curve itself degenerates. Section~\ref{sec:small-chi} examines the resulting low order coefficient behavior.

\subsection{Polynomial WKB recursion}
\label{sec:radial-recursion}
Instead of expanding $\widehat Q$ directly, we use the ansatz
\begin{equation}
 \phi(u)=\exp\left[\frac{i}{\hbar}\int^u\frac{P(u';\hbar)}{F(u')}\,du'\right].
 \label{eq:polynomial-ansatz}
\end{equation}
Substitution into~\eqref{eq:poly-ODE} gives an exactly equivalent Riccati equation,
\begin{equation}
 P^2-i\hbar DP=Q,\qquad
 P=\sum_{j\ge0}\hbar^jP_j,
 \qquad P_0=\frac Y2.
 \label{eq:polynomial-Riccati}
\end{equation}
The coefficients $Q_j$ in this expansion are polynomials. To see how the angular calculation enters, rewrite its output using $C=i\chi\Omega/2$:
\begin{equation}
 A(c)=A_{-1}\Omega+A_0+\sum_{j\ge1}A_j\Omega^{-j},\qquad
 A_{-1}=\frac{i\chi\eta}{2},\quad
 A_j=\left(\frac2{i\chi}\right)^j\mathcal A_j^{\rm nor}\quad(j\ge0).
 \label{eq:angular-radial-conversion}
\end{equation}
Inserting it into the radial potential, we obtain
\begin{align}
 Q_1&=u(u-1)[2\chi m+(A_{-1}-\chi m)u],\nonumber\\
 Q_2&=u^2(u-1)[A_0+s(s+1)-(s^2-1)u],\nonumber\\
 Q_j&=u^2(u-1)A_{j-2},\qquad j\ge3.
 \label{eq:radial-Qj}
\end{align}
Together with~\eqref{eq:G}, these are the complete inputs to the recurrence
\begin{equation}
 P_j=\frac1{2P_0}\left(Q_j+i DP_{j-1}
                  -\sum_{r=1}^{j-1}P_rP_{j-r}\right),\qquad j\ge1.
 \label{eq:Riccati-recursion}
\end{equation}

The recursion~\eqref{eq:Riccati-recursion} starts from $P_0=Y/2$. Since $Y$ enters only through $P_0$ and $Y^2=G_\sigma$, each $P_j$ has a unique decomposition $P_j=P_j^{\rm e}+P_j^{\rm o}$, where $P_j^{\rm e}$ and $P_j^{\rm o}/Y$ are rational functions of $u$. We call them the branch even part and the branch odd part. They are even and odd under the sheet exchange $Y\to-Y$, which maps the solution started at $P_0=Y/2$ to the second formal solution started at $-Y/2$. The two branches are therefore
\begin{equation}
 P^{(\pm)}=P_{\rm even}\pm P_{\rm odd},\qquad
 P_{\rm even}=\sum_{j\ge1}\hbar^jP_j^{\rm e},\qquad
 P_{\rm odd}=\sum_{j\ge0}\hbar^jP_j^{\rm o},\qquad
 P_0^{\rm o}=\frac Y2,
 \label{eq:branch-parity}
\end{equation}
and $P=P^{(+)}$. The labels ${\rm e}$ and ${\rm o}$ refer to this sheet exchange, not to even and odd powers of $\hbar$. This distinction matters because $Q(u;\hbar)$ contains odd powers of $\hbar$. Subtracting the Riccati equations for $P^{(+)}$ and $P^{(-)}$ gives $P_{\rm even}=(i\hbar/2)\,D\log P_{\rm odd}$. Therefore, for any closed cycle $\gamma$ on the spectral curve, we define
\begin{equation}
 \frac{P_{\rm odd}}F\,du=\sum_{j\ge0}\hbar^j\lambda_j,
 \qquad \Pi_j^\gamma=\oint_\gamma\lambda_j.
 \label{eq:quantum-periods}
\end{equation}
If $\Psi=\exp[(i/\hbar)\int\widehat P\,du]$, its normal form momentum is
\begin{equation}
 \widehat P=\frac PF-\frac{i\hbar}{2}\frac{F'}F,
 \qquad \widehat P_{\rm odd}\,du=\frac{P_{\rm odd}}F\,du.
 \label{eq:gauge-invariance}
\end{equation}
Since the gauge shift is branch even, the polynomial form and the normal form give the same periods.

For example, the first recurrence step gives
\begin{equation}
 P_1^{\rm e}=\frac{i FG'}{4G},\qquad P_1^{\rm o}=\frac{Q_1Y}{G},
 \qquad G=G_\sigma.
 \label{eq:P1}
\end{equation}
At the next step,
\begin{align}
 P_2^{\rm e}&=i F\partial_u(Q_1/G),\nonumber\\
 P_2^{\rm o}&=\frac YG\left[Q_2+i F\partial_u\left(\frac{i FG'}{4G}\right)
                 -\left(\frac{i FG'}{4G}\right)^2-\frac{Q_1^2}{G}\right].
 \label{eq:P2}
\end{align}
Both parities must be propagated: $P_j^{\rm e}$ does not contribute directly to $\Pi_j^\gamma$, but contributes to the next Riccati step. The factors in~\eqref{eq:radial-Qj} imply that the numerator of $P_j^{\rm o}$ contains $F$ for $j\ge2$. Hence
\begin{equation}
 \lambda_j=R_j(u)\frac{du}{Y^{2k_j+1}},\qquad
 k_j=\left\lfloor\frac{3j}{2}\right\rfloor-1,\qquad j\ge2.
 \label{eq:raw-form}
\end{equation}
Here $R_j$ is a polynomial and $\lfloor\cdot\rfloor$ is the floor function. The explicit implementation and a worked example are given in Appendix~\ref{app:polynomial}. Note that we always use the unreduced $P_j$ in~\eqref{eq:Riccati-recursion}. Exact terms can be dropped only when we compute the periods.

\subsection{Why every higher period reduces to two elliptic periods}
\label{sec:period-reduction}
The raw forms~\eqref{eq:raw-form} may have higher poles at the turning points, but they have no new poles at the punctures. Let us first see this locally. At $u=0$,
\begin{equation}
 Q_0(0)=1-\sigma\ne0,\quad Q_1=O(u),\quad Q_j=O(u^2)\ (j\ge2),
 \quad D=O(u^2)\partial_u.
\end{equation}
Induction in~\eqref{eq:Riccati-recursion} gives $P_1=O(u)$ and $P_j=O(u^2)$ for $j\ge2$. At $u=1$, all $Q_j$ with $j\ge1$ vanish at least linearly and $D=O(u-1)\partial_u$, so $P_j=O(u-1)$. Division by $F=u^2(u-1)$ therefore leaves $\lambda_j$, $j\ge2$, regular at both physical punctures.

At a simple zero $u_*$ of $G$, use $Y$ as the local coordinate. Since $u-u_*=O(Y^2)$ and $du=O(Y)\,dY$, every monomial in~\eqref{eq:raw-form} has an expansion in even powers of $Y$ times $dY$, with no $dY/Y$ term. At infinity, using $u^{-1/2}$ itself as a local coordinate, $u^pdu/Y^{2k+1}$ is $(u^{-1/2})^{6k-2p}$ times a power series in $u^{-1}$, multiplied by $d(u^{-1/2})$. The exponent is even at every term, again excluding a residue. Therefore the higher forms are differentials of the second kind on the compact nonsingular curve.

On a genus-one curve, the space of second kind forms modulo exact meromorphic differentials has dimension two. A convenient basis is $du/Y$ and $u\,du/Y$, so
\begin{equation}
 \lambda_j\equiv(\mathcal C_{j,0}+\mathcal C_{j,1}u)\frac{du}{Y}
                       \pmod{\text{exact}},\qquad j\ge2,
 \label{eq:two-period-revised}
\end{equation}
which implies
\begin{equation}
 \Pi_j^{\gamma_A}=\mathcal C_{j,0}J_0+\mathcal C_{j,1}J_1,
 \qquad J_0=\oint_{\gamma_A}\frac{du}{Y},\quad J_1=\oint_{\gamma_A}\frac{u\,du}{Y}.
 \label{eq:two-period-theorem}
\end{equation}
The coefficients $\mathcal C_{j,0}$ and $\mathcal C_{j,1}$ depend on the mode parameters and rotation, but not on $u$. Note that $J_0$ and $J_1$ are the integrals of different differentials on the same cycle, and should not be confused with classical actions on two independent cycles. This reduction is a genus-one instance of the Griffiths--Hermite reduction~\cite{Griffiths}.

The first two orders are exceptional. The leading differential is meromorphic at both physical punctures, while $\lambda_1$ still has a puncture pole at $u=0$. Only for $j\ge2$ are the WKB differentials regular at both punctures and hence reducible, modulo exact forms, to the two elliptic basis periods $J_0$ and $J_1$. Define
\begin{equation}
 J_{-1}=\oint_{\gamma_A}\frac{du}{uY},\qquad J_h=\oint_{\gamma_A}\frac{du}{(u-1)Y}.
\end{equation}
Reduction of $\lambda_0$ modulo an exact form, and direct use of $P_1^{\rm o}$, give
\begin{align}
 \Pi_0^{\gamma_A}&=-\frac\sigma4J_1+\frac\sigma2J_0-2J_{-1}+2J_h,\nonumber\\
 \Pi_1^{\gamma_A}&=\frac\chi2[(i\eta-2m)J_0+4mJ_{-1}].
 \label{eq:low-periods}
\end{align}
The distinction between $\gamma_A$ and $\gamma_{\rm out}$ affects the leading period at the two points of the spectral curve above $u=1$. We denote them by $p_1^{(\pm)}=(u,Y)=(1,\pm2)$. Since $u=1$ is not a branch point, $u-1$ is a local coordinate at either point, and the residues in that coordinate are
\begin{equation}
 \mathop{\mathrm{Res}}\limits_{p_1^{(+)}}\lambda_0=+1,\qquad
 \mathop{\mathrm{Res}}\limits_{p_1^{(-)}}\lambda_0=-1.
 \label{eq:horizon-residues}
\end{equation}
Writing $\Pi_j:=\Pi_j^{\gamma_{\rm out}}$ for the periods along the outer cycle, the signed puncture loops in~\eqref{eq:outer-cycle} give $2\pi i-(-2\pi i)=4\pi i$, and therefore
\begin{equation}
 \Pi_0=\Pi_0^{\gamma_A}+4\pi i,\qquad
 \Pi_j=\Pi_j^{\gamma_A}\quad(j\ge1).
 \label{eq:outer-residue}
\end{equation}
The second equality follows because all higher relevant differentials, including $\lambda_1$, are regular at the two lifts of $u=1$. Once the homology class is fixed,~\eqref{eq:outer-residue} is an exact residue identity. At leading order, it corresponds to the relation between the two contours $C_{t,o}$ and $C_{t,i}$ in~\cite{KH}, which connect the turning points outside and inside the event horizon, respectively.

For explicit elliptic evaluation, set
\begin{equation}
 \mathsf m=\frac{u_--u_+}{u_--u_0},\quad
 \mathsf n_0=\frac{u_--u_+}{u_-},\quad
 \mathsf n_h=\frac{u_--u_+}{u_--1},\quad
 \mathcal N_{\rm ell}=\frac4{i\chi\sqrt{u_--u_0}}.
\end{equation}
We use the elliptic parameter convention $\mathsf m=k^2$, rather than the modulus $k$. We define
\begin{align}
\Kell(\mathsf m)
 &:=\int_0^{\pi/2}\frac{d\theta}{\sqrt{1-\mathsf m\sin^2\theta}}, \qquad
 \Eell(\mathsf m)
 :=\int_0^{\pi/2}\sqrt{1-\mathsf m\sin^2\theta}\,d\theta, \nonumber\\
\Piell(\mathsf n\vert\mathsf m)
 &:=\int_0^{\pi/2}
 \frac{d\theta}{(1-\mathsf n\sin^2\theta)\sqrt{1-\mathsf m\sin^2\theta}}. 
 \label{eq:Pi-definition}
\end{align}
The square root in $\mathcal N_{\rm ell}$ has positive real part at the reference spin and is continued with the cycle. Substitution $u=u_--(u_--u_+)\sin^2\theta$ in the lifted cut integrals gives
\begin{align}
 J_0&=\mathcal N_{\rm ell}\Kell(\mathsf m),& J_1&=\mathcal N_{\rm ell}[u_0\Kell(\mathsf m)+(u_--u_0)\Eell(\mathsf m)],\nonumber\\
 J_{-1}&=\frac{\mathcal N_{\rm ell}}{u_-}\Piell(\mathsf n_0\vert\mathsf m),&
 J_h&=\frac{\mathcal N_{\rm ell}}{u_--1}\Piell(\mathsf n_h\vert\mathsf m).
 \label{eq:elliptic-period-basis}
\end{align}
These integral definitions also fix the analytic continuation convention.

The reduction can be implemented recursively using only the physical cubic $Y^2=G_\sigma(u)$. We transport $\gamma_A$ continuously with the branch points as $\sigma$ varies. Define
\begin{equation}
 I_{p,k}=\oint_{\gamma_A}\frac{u^p\,du}{Y^{2k+1}},\qquad p,k\ge0,
 \qquad I_{0,0}=J_0,\quad I_{1,0}=J_1.
 \label{eq:moments}
\end{equation}
The recurrence relations follow from integrating exact differentials over the closed cycle. For $k=0$,
\begin{equation}
 0=\oint_{\gamma_A} d(u^pY).
 \label{eq:exact-diff-k0}
\end{equation}
Using $Y'=G_\sigma'/(2Y)$ and the explicit cubic gives
\begin{equation}
 I_{p+2,0}=\frac{10(p+1)}{2p+3}I_{p+1,0}
 -\frac{8(2p+1)}{2p+3}I_{p,0}
 -\frac{8p(1-\sigma)}{\sigma(2p+3)}I_{p-1,0}.
 \label{eq:moment-row0}
\end{equation}
The last term vanishes for $p=0$. In particular,
\begin{equation}
 I_{2,0}=\frac{10}{3}J_1-\frac{8}{3}J_0.
 \label{eq:I20-example}
\end{equation}
For $k\ge1$, the analogous identity
\begin{equation}
 0=\oint_{\gamma_A} d\!\left(\frac{u^p}{Y^{2k-1}}\right)
 \label{eq:exact-diff-k}
\end{equation}
gives
\begin{equation}
 I_{p+2,k}=\frac{10}{3}I_{p+1,k}-\frac83I_{p,k}
 +\frac{2p}{3\sigma(2k-1)}I_{p-1,k-1}.
 \label{eq:moment-rowk}
\end{equation}
Thus, at fixed $k$, the power of $u$ can first be reduced to $p=0,1$. The remaining denominator power is lowered with the curve equation itself: $Y^2=G_\sigma(u)$ relates $I_{p,k-1}$ to a linear combination of $I_{p,k},\ldots,I_{p+3,k}$. After the $p$ reduction, the first two entries at each pole order satisfy
\begin{equation}
 \binom{I_{0,k}}{I_{1,k}}=M_k(\sigma)\binom{I_{0,k-1}}{I_{1,k-1}},
\end{equation}
where
\begin{equation}
 M_k(\sigma)=\frac1{(2k-1)(\sigma+27)}
 \begin{pmatrix}
 \dfrac{-4k\sigma+54k+5\sigma-45}{4}&\dfrac{\sigma(6k-7)}8\\[2mm]
 \dfrac{-4k\sigma+5\sigma+15}{2}&\dfrac{6k\sigma+54k-7\sigma-63}{4}
 \end{pmatrix}.
 \label{eq:moment-matrix}
\end{equation}
Repeated application therefore reduces every $I_{p,k}$ to $I_{0,0}=J_0$ and $I_{1,0}=J_1$. If $R_j(u)=\sum_pr_{j,p}u^p$, then $\Pi_j^{\gamma_A}=\sum_pr_{j,p}I_{p,k_j}$. The same transported cycle convention yields the Gauss--Manin system
\begin{equation}
 \frac d{d\sigma}\binom{J_0}{J_1}=
 \begin{pmatrix}
 -\dfrac9{\sigma(\sigma+27)}&-\dfrac1{4(\sigma+27)}\\[2mm]
 \dfrac{\sigma+15}{\sigma(\sigma+27)}&-\dfrac{\sigma+18}{\sigma(\sigma+27)}
 \end{pmatrix}\binom{J_0}{J_1}.
 \label{eq:Gauss-Manin}
\end{equation}

Having completed the general reduction, we now give the first nontrivial higher period as an explicit example. With $q=\bar L(\bar L+1)$ and $\eta=2\bar L+1$, the general spin result is
\begin{align}
 \Pi_2^{\gamma_A}&=\mathcal C_{2,0}J_0+\mathcal C_{2,1}J_1,\nonumber\\
 \mathcal C_{2,0}&=2s^2+
 \frac{6(2\sigma^2+49\sigma+9)q
 +12(\sigma+9)(m^2+2i m\eta)
 +2\sigma^2+37\sigma-171}{12(\sigma+27)},\nonumber\\
 \mathcal C_{2,1}&=-s^2-
 \frac{6\sigma[(\sigma+23)q+m^2+2i m\eta]
 +\sigma^2+17\sigma-54}{12(\sigma+27)}.
 \label{eq:Pi2-general-spin}
\end{align}
At fixed $(\bar L,m,\chi)$, $\Pi_0^{\gamma_A}$ and $\Pi_1^{\gamma_A}$ are independent of $s$, while explicit spin dependence first appears in $\Pi_2^{\gamma_A}$.

\subsection{Formal quantization and frequency inversion}
\label{sec:frequency-inversion}
We use the all-order extension of the quantization condition in~\cite{KH,KN}:
\begin{equation}
 \mathcal Q(\Omega)=\Pi_0\Omega+\Pi_1+\frac{\Pi_2}{\Omega}
              +\frac{\Pi_3}{\Omega^2}+\cdots
       =2\pi\left(N+\frac12\right),
 \label{eq:normal-quantization}
\end{equation}
The direct numerical comparisons in Section~\ref{sec:normal-validation} support this condition at all the orders tested there.

Since $\Pi_0\neq0$, the quantization condition~\eqref{eq:normal-quantization} can be solved for $\Omega$ as a series in $1/(N+1/2)$, and the coefficients in~\eqref{eq:intro-series} are uniquely determined. The first ones are
\begin{align}
 d_{-1}&=\frac{2\pi}{\Pi_0},\qquad d_0=-\frac{\Pi_1}{\Pi_0},\qquad
 d_1=-\frac{\Pi_2}{2\pi},\nonumber\\
 d_2&=-\frac{\Pi_1\Pi_2+\Pi_0\Pi_3}{4\pi^2},\nonumber\\
 d_3&=-\frac{\Pi_0^2\Pi_4+2\Pi_0\Pi_1\Pi_3+\Pi_0\Pi_2^2+\Pi_1^2\Pi_2}{8\pi^3}.
 \label{eq:d-low}
\end{align}
In this way, we can compute the coefficients to any finite order. To obtain $d_J$ ($J\ge1$), we proceed in four steps:
\begin{enumerate}[label=(\arabic*)]
 \item Compute the angular coefficients through $\mathcal A^{\rm nor}_{J-1}$ by the recursion~\eqref{eq:hermite-recursion}, which requires the solvability condition at $\varepsilon^{2J}$.
 \item Insert them into~\eqref{eq:radial-Qj}, and compute $\lambda_j$ through $j=J+1$ by~\eqref{eq:Riccati-recursion}.
 \item Reduce the periods to $J_0$ and $J_1$ by the moment relations, and evaluate them by~\eqref{eq:elliptic-period-basis}.
 \item Invert the quantization condition~\eqref{eq:normal-quantization} order by order.
\end{enumerate}
Note that the period order, the angular order and the frequency order are different indices.

At fixed order, the period coefficients are analytic in $0<\chi<1$ along a transported cycle, away from zeros of $\Pi_0$. As shown in the next subsection, however, the expansion is not uniform near $\chi=0$.

\subsection{Rotation dependence and the small spin limit}
\label{sec:small-chi}
\begin{figure}[!tbp]
 \centering
 \includegraphics[width=.49\textwidth]{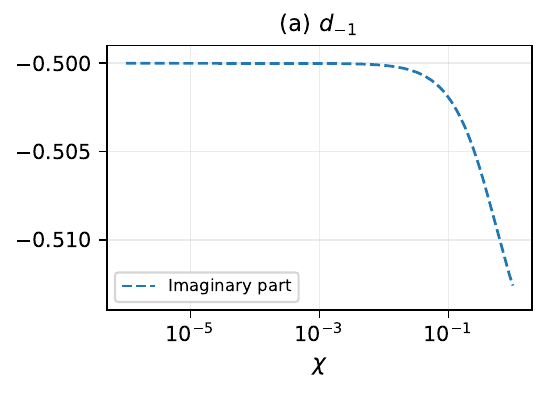}\hfill
 \includegraphics[width=.49\textwidth]{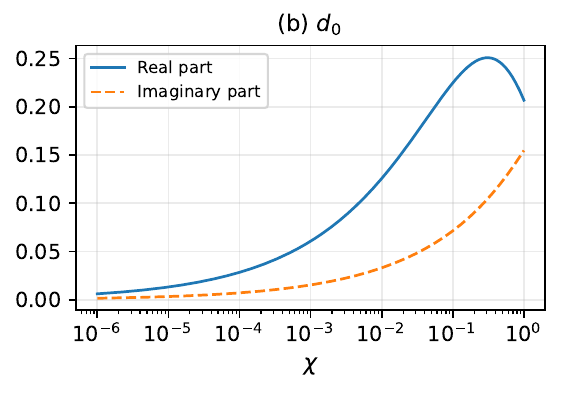}\\
 \includegraphics[width=.49\textwidth]{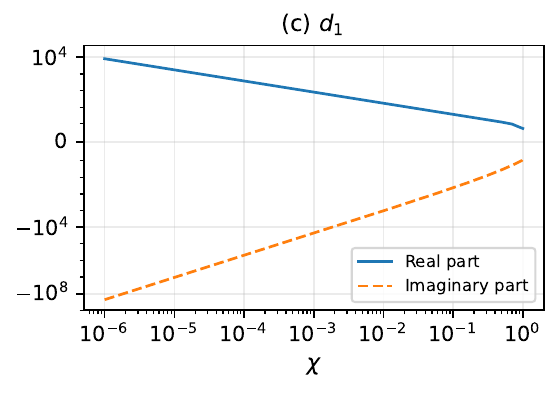}\hfill
 \includegraphics[width=.49\textwidth]{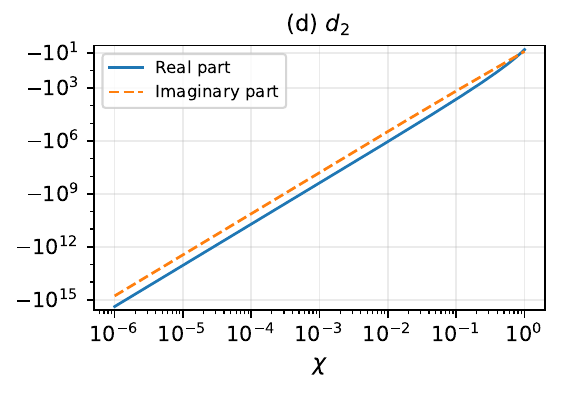}
 \caption{The coefficients $d_{-1}$, $d_0$, $d_1$ and $d_2$ as functions of $\chi$ for $(s,m,\bar L)=(-2,-2,0)$. Solid and dashed curves show $\operatorname{Re}d_j$ and $\operatorname{Im}d_j$, respectively ($\operatorname{Re}d_{-1}=0$ is not plotted). The horizontal axis is logarithmic, and panels (c) and (d) use a signed logarithmic scale with a linear region for $|d|<1$. As $\chi\to0$, $d_1$ and $d_2$ diverge as predicted by~\eqref{eq:smallchi-d12}, while $d_{-1}$ and $d_0$ remain finite. The numerical details are given in Appendix~\ref{app:normal-numerics}.}
 \label{fig:rotation}
\end{figure}

Let us now look at the rotation dependence of the coefficients $d_j(\chi)$. Figure~\ref{fig:rotation} shows the first four coefficients for the gravitational input $(s,m,\bar L)=(-2,-2,0)$. The coefficients $d_{-1}$ and $d_0$ remain finite at small spin, while $d_1$ and $d_2$ diverge. As we will show below, their fractional powers can be derived from the curve.\footnote{The non-uniform small angular momentum limit of highly damped Kerr modes was investigated in~\cite{DGM}. The fractional powers in~\eqref{eq:smallchi-d12} are derived here independently from the degeneration of the curve.}

To study the compact turning points as $\chi\to0$, we set $\epsilon_\chi=\chi^{2/3}$ and $u=v/\epsilon_\chi$. The curve becomes
\begin{equation}
 Y^2=v^3+4-5\epsilon_\chi v^2+8\epsilon_\chi^2v-4\epsilon_\chi^3
       \quad \longrightarrow \quad Y_0^2=v^3+4.
 \label{eq:smallchi-curve}
\end{equation}
On the limiting transported compact cycle $\gamma_A^{(0)}$, define positive real constants
\begin{align}
 B_j&=i\oint_{\gamma_A^{(0)}}\frac{v^j\,dv}{\sqrt{v^3+4}},\qquad j=0,1,\nonumber\\
 B_0&=\frac{2^{1/3}\,\Gamma(1/3)^3}{2\pi}=3.855242593319996262\ldots,\nonumber\\
 B_1&=\frac{8\cdot2^{2/3}\pi^2}{\sqrt3\,\Gamma(1/3)^3}=3.763808518565342108\ldots.
 \label{eq:smallchi-periods}
\end{align}
The change of variables immediately gives
\begin{equation}
 J_0\sim-i B_0\chi^{-2/3},\qquad
 J_1\sim-i B_1\chi^{-4/3}.
 \label{eq:smallchi-J}
\end{equation}
On the compact cycle the leading differential is of order $\epsilon_\chi^2$, whereas its horizon loop contribution is unchanged. Hence
\begin{equation}
 \Pi_0=4\pi i+O(\chi^{4/3}),\qquad \Pi_1=O(\chi^{1/3}),\qquad
 d_{-1}=-\frac{i}{2}+O(\chi^{4/3}),\quad d_0=O(\chi^{1/3}).
 \label{eq:smallchi-affine}
\end{equation}
In the second statement, the term $\chi J_0$ in~\eqref{eq:low-periods} dominates. The term $\chi J_{-1}$ is smaller. At leading order this gives $d_0\sim2^{1/3}\Gamma(1/3)^3(i\eta-2m)\chi^{1/3}/(16\pi^2)$, so that $\operatorname{Re}d_0\sim-2^{1/3}m\,\Gamma(1/3)^3\chi^{1/3}/(8\pi^2)$. The vanishing of the real part as $\chi^{1/3}$ agrees with the small spin behavior of the leading real frequency found in~\cite{KH}. These limits would be incorrect if the $4\pi i$ outer cycle shift were omitted.

The small spin limit of the fixed spin coefficients in~\eqref{eq:smallchi-affine} is therefore not the strict Schwarzschild high overtone expansion: in particular $d_0\to0$ here, whereas the familiar Schwarzschild asymptotics has a non-zero real constant term. This shows that the limits $N\to\infty$ and $\chi\to0$ do not commute. Since $|c|\sim\chi N/4$ at leading order, the elementary large angular parameter requirement is $N\gg\chi^{-1}$. The radial coefficients derived below impose a stronger condition.

For the next two periods, the polynomial reduction of Section~\ref{sec:period-reduction} gives, for this gravitational input,
\begin{equation}
 \mathcal C_{2,0}\longrightarrow\frac{317}{36}-\frac43i,
 \quad \mathcal C_{2,1}\longrightarrow-\frac{23}{6},
 \quad \chi\mathcal C_{3,1}\longrightarrow\frac{46}{9}(4+i).
 \label{eq:smallchi-C}
\end{equation}
The $\mathcal C_{3,0}J_0$ term is subleading to the last contribution. Inserting~\eqref{eq:smallchi-J}--\eqref{eq:smallchi-C} into~\eqref{eq:d-low} yields
\begin{align}
 \operatorname{Re}d_1&\sim\frac{2^{1/3}\Gamma(1/3)^3}{3\pi^2}\chi^{-2/3}
                  =0.8181078874\,\chi^{-2/3},\nonumber\\
 \operatorname{Im}d_1&\sim-\frac{46\cdot2^{2/3}\pi}{3\sqrt3\,\Gamma(1/3)^3}\chi^{-4/3}
                  =-2.2962768006\,\chi^{-4/3},\nonumber\\
 d_2&\sim-\frac{368\cdot2^{2/3}\pi}{9\sqrt3\,\Gamma(1/3)^3}(4+i)\chi^{-7/3}
       =(-24.49361921-6.123404801i)\chi^{-7/3}.
 \label{eq:smallchi-d12}
\end{align}
$\operatorname{Re}d_1$ and $\operatorname{Im}d_1$ begin at different powers. The product $\Pi_1\Pi_2$ is subleading in $d_2$, so its leading singularity follows from $\Pi_0\Pi_3$ alone.

By~\eqref{eq:smallchi-d12}, the first inverse overtone correction $d_1/(N+\frac12)$ becomes of order one when $N\sim\chi^{-4/3}$, whereas $d_2/(N+\frac12)^2$ is then still suppressed by $\chi^{1/3}$. The scale reflects the compact turning points at $|u|\sim\chi^{-2/3}$ in~\eqref{eq:smallchi-curve}. For this gravitational input the fixed spin expansion is therefore controlled only for $N\gg\chi^{-4/3}$, which is more restrictive than $N\gg\chi^{-1}$ at small spin. The same scale follows from the operator $D$. With $u=v\chi^{-2/3}$, it becomes $D=\chi^{-4/3}v^2(v-\chi^{2/3})\partial_v$, so that the local WKB parameter is $\hbar\chi^{-4/3}$. In fact, with $\tau=\chi^{1/3}$ and $u=v/\tau^2$, the rescaled potentials $\tau^{4j}Q_j$ are polynomials in $\tau$, and $du/F=\tau^4dv/[v^2(v-\tau^2)]$. Since the limiting curve $Y_0^2=v^3+4$ is nondegenerate, the compact periods satisfy $\Pi_r^{\gamma_A}=O(\chi^{(4-4r)/3})$, and the inversion of~\eqref{eq:normal-quantization} gives $d_j=O(\chi^{-4j/3})$ for every fixed $j\ge0$. More generally, we have found that $d_j\sim h_j\chi^{p_j}$ with non-zero constants $h_j$ and $p_j=(-4j+K_j)/3$, where $K_j=0,1,2,3,0,1$ for $j\equiv1,2,3,4,5,0\pmod 6$, up to $j=60$ for several families. This is consistent with the bound, which is saturated for $j\equiv1,5\pmod6$. The bound identifies $\chi^{-4/3}/(N+\frac12)$ as the local expansion parameter at fixed order, but it is not uniform in $j$, since the coefficients grow factorially. The transition between the Kerr and Schwarzschild asymptotic spectra near $|a\omega|\approx1$ was discussed in~\cite{DGM}. We do not analyze the transition region in this paper.

We have checked the constants in~\eqref{eq:smallchi-periods} by direct numerical integration, and the approach of $d_1$ and $d_2$ to~\eqref{eq:smallchi-d12} by rescaling them with the predicted powers.
\section{Numerical and analytic tests of the expansion}
\label{sec:normal-checks}
In this section, we test the expansion obtained above in two ways. We first compare its finite truncations with QNM frequencies obtained by solving the coupled angular and radial equations directly. We then compare the scalar coefficients with those obtained from the exact quantization condition for extremal Kerr black holes.

\subsection{Comparison with direct numerical solutions}
\label{sec:normal-validation}
Let us now test the expansion~\eqref{eq:intro-series} against QNM frequencies obtained by solving the angular and radial equations directly. At each trial frequency, we compute the angular eigenvalue numerically at $c=\chi\Omega/2$ on the transported branch, and insert it into the radial continued fraction (see Appendix~\ref{app:direct}). The large-$c$ series of the angular eigenvalue is not used here. We consider the modes with
\begin{equation}
 (s,\ell,m,\bar L)=(0,2,-2,0),\quad(-1,2,-2,0),\quad(-2,2,-2,0).
 \label{eq:common-normal-benchmarks}
\end{equation}
For each of them, we choose $N=40,60,80,100,120$ at $\chi=1/2,2/3,4/5$, and $N=80,100,120$ at $\chi=9/10,19/20$. We label the roots by $N$. Their overtone numbers are discussed in Appendix~\ref{app:normal-numerics}.

\begin{figure*}[!t]
 \centering
 \includegraphics[width=\textwidth]{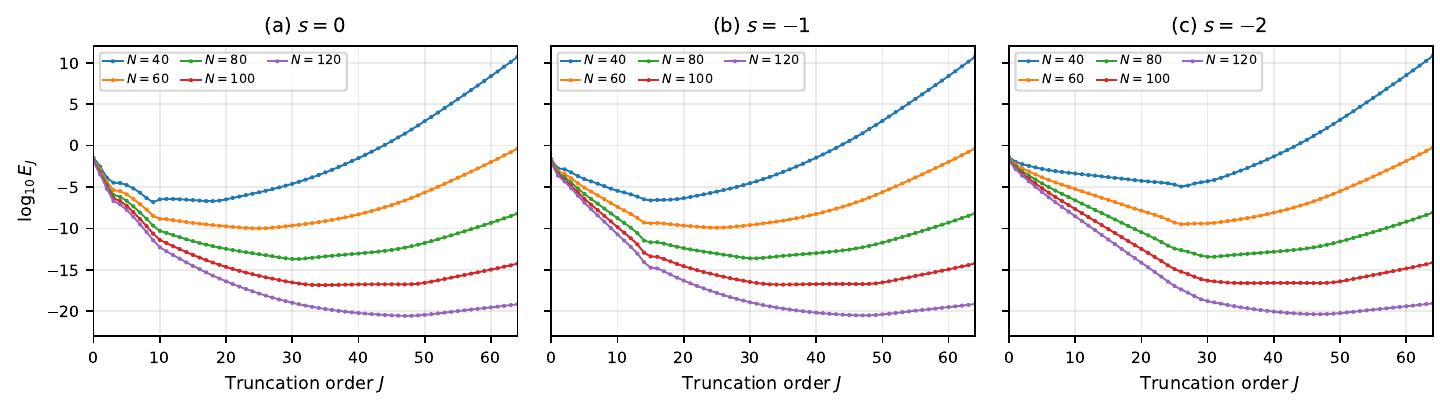}
 \caption{The truncation error $E_J$ at $\chi=4/5$ for (a)~$s=0$, (b)~$s=-1$ and (c)~$s=-2$, with $(\ell,m,\bar L)=(2,-2,0)$ and $N=40,60,80,100,120$. No resummation is used. The numerical uncertainties of the direct roots are at most $4\times10^{-32}$, and are negligible in this plot.}
 \label{fig:three-spins}
\end{figure*}

We compare each root with the truncated expansion
\begin{equation}
 \Omega^{[J]}(N)=d_{-1}g^{-1}+d_0+\sum_{j=1}^{J}d_jg^j,
       \qquad g=(N+1/2)^{-1},
 \label{eq:truncation}
\end{equation}
and define the error $E_J=|\Omega^{\rm direct}(N)-\Omega^{[J]}(N)|$ for $0\le J\le J_{\rm scan}$, where $J_{\rm scan}=159$. We denote by $J_{\rm best}$ the order that minimizes $E_J$, and by $J_{\rm LT}$ the order of the smallest term $|d_jg^j|$ with $1\le j\le J_{\rm scan}$. The latter is determined by the coefficients alone. Representative results are shown in Table~\ref{tab:normal-truncation} and Figure~\ref{fig:three-spins}.
\begin{table}[!tbp]
 \centering
 \caption{Tests of the truncated expansion~\eqref{eq:truncation} for $(\ell,m,\bar L)=(2,-2,0)$. Here $J_{\rm best}$ is the order that minimizes the error $E_J$, and $J_{\rm LT}$ is the order of the smallest term $|d_jg^j|$.}
 \label{tab:normal-truncation}
 \begin{tabular}{ccrrrrr}
 \toprule
 $s$&$\chi$&$N$&$J_{\rm scan}$&$J_{\rm best}$&$J_{\rm LT}$&$E_{J_{\rm best}}$\\\midrule
 $0$&$1/2$&100&159&23&23&$8.980\times10^{-11}$\\
 $0$&$4/5$&120&159&47&47&$2.706\times10^{-21}$\\
 $-1$&$1/2$&40&159&14&15&$5.273\times10^{-4}$\\
 $-1$&$4/5$&120&159&47&47&$3.020\times10^{-21}$\\
 $-2$&$1/2$&40&159&2&5&$2.679\times10^{-2}$\\
 $-2$&$4/5$&80&159&30&31&$3.704\times10^{-14}$\\
 $-2$&$4/5$&120&159&47&47&$4.254\times10^{-21}$\\
 $-2$&$19/20$&120&159&48&53&$3.674\times10^{-25}$\\
 \bottomrule
 \end{tabular}
\end{table}

We find that the expansion works very well. At $\chi=4/5$ and $N=120$, the error reaches $10^{-21}$ around $J=47$ for all three fields, without any resummation. As expected for an asymptotic series, the error first decreases, and then increases beyond the optimal order. In fact, our high-order data show that $d_j$ grows factorially, and $J_{\rm LT}$ grows roughly in proportion to $N$. Note that $J_{\rm best}$ and $J_{\rm LT}$ need not coincide, and the error does not always decrease monotonically, because of the interference among complex terms. When $N$ and $\chi$ are small, the expansion does not work so well. For instance, the gravitational mode at $\chi=1/2$ and $N=40$ has $|c|\simeq5.1$, and its best error is about $3\times10^{-2}$. We have also checked that the direct roots are stable under the variation of the numerical parameters, and that the standard and dual radial formulations give the same roots (see Appendix~\ref{app:normal-numerics}).
\subsection{An independent extremal scalar comparison}
\label{sec:HS}
In~\cite{HS}, exact WKB analysis was applied to the scalar QNMs of extremal Kerr and extremal Reissner--Nordstr\"om black holes. For extremal Kerr black holes ($a=M$), the two horizons coalesce, and the radial Teukolsky equation becomes a doubly confluent Heun equation. The QNM boundary condition was formulated as an exact quantization condition in terms of a $B$ cycle quantum period, the WKB periods were computed to high orders, and the quantization condition, resummed by the Borel--Pad\'e method, was tested against high-precision numerical QNM frequencies. The high overtone expansion of this quantization condition was not discussed there, and we carry it out in this paper.

This gives an independent check of the coefficients $d_j(\chi)$ obtained above, which have finite limits as $\chi\to1^-$ for the scalar sectors considered below. The exact quantization condition (EQC) in~\cite{HS} reads
\begin{equation}
 \mathcal S\!\left[\Pi_B(\mathcal A,\mu;\hbar_{\rm ext})\right]
       =2\pi\hbar_{\rm ext}\left(n_{\rm ext}+\frac12\right),\qquad
 \Pi_B=\sum_{k\ge0}\hbar_{\rm ext}^{2k}\Pi_B^{(k)}(\mathcal A,\mu),
 \label{eq:HS-EQC}
\end{equation}
where $\mathcal S$ denotes the Borel resummation. The $B$ cycle is the cycle $\gamma_{\beta_-\beta_+}$ in~\cite{HS}, obtained by lifting the arc joining the two turning points $\beta_\pm$ and returning on the second sheet. The integer $n_{\rm ext}$ labels the extremal problem. The Stokes geometry was checked in~\cite{HS} at representative parameters. As in~\cite{HS}, we assume that it does not change in the high overtone regime. In our normalization, the dictionary is
\begin{align}
 \hbar_{\rm ext}&=\frac{i\sqrt2}{\sqrt{\Omega(\Omega-m)}},\qquad
 \mu=\sqrt{\frac{2\Omega}{\Omega-m}},\nonumber\\
 \mathcal A^2&=\frac{2[A(\Omega/2)+\Omega^2/4+1/4-m\Omega]}{\Omega(\Omega-m)}.
 \label{eq:HS-dictionary}
\end{align}
The quantity $\lambda_{\ell m}$ in~\cite{HS} corresponds to $A(c)+c^2$ here, where $A(c)$ is the separation constant for $s=0$. This shift is essential, because the normal $A(c)$ is linear at leading order, whereas $c^2$ is quadratic.

The large $|\Omega|$ counting now follows directly from these definitions. Since the normal $A(c)$ grows only linearly in $\Omega$, the dictionary~\eqref{eq:HS-dictionary} gives on the chosen branches
\begin{equation}
 \hbar_{\rm ext}=O(\Omega^{-1}),\qquad
 (\mathcal A,\mu)\longrightarrow\left(\frac1{\sqrt2},\sqrt2\right).
 \label{eq:HS-order-counting}
\end{equation}
We write $\Pi_{B,*}^{(k)}:=\Pi_B^{(k)}(1/\sqrt2,\sqrt2)$ for the period coefficients at this limiting point, so that $\Pi_B^{(k)}=\Pi_{B,*}^{(k)}+O(\Omega^{-1})$. Therefore the $k$th WKB contribution starts at
\begin{equation}
 \hbar_{\rm ext}^{2k}\Pi_B^{(k)}=O(\Omega^{-2k}).
 \label{eq:HS-WKB-order}
\end{equation}
This determines how many quantum period terms can contribute to a prescribed inverse overtone order.

At the limiting point, the classical $B$ period in~\cite{HS} is given by an explicit integral. In the rescaled extremal coordinate $\xi$ of Appendix~\ref{app:HS-bridge},
\begin{equation}
 \Pi_{B,*}^{(0)}=2i\int_{\beta_-}^{\beta_+}\frac{R_H(\xi)}{\xi^2}\,d\xi,
 \qquad
 R_H^2=(\xi+\sqrt2)(\xi+1/\sqrt2)(\xi^2+\xi/\sqrt2+1),
 \label{eq:HS-B-integral}
\end{equation}
where $\beta_\pm=(-1\pm i\sqrt7)/(2\sqrt2)$ are the two turning points. The path intersects the positive real axis of $\xi$, that is, to the right of the double pole at $\xi=0$, and $R_H$ is the branch with $R_H(0)=+1$. Moving the path to the other side of the pole changes the value by $4\pi\sqrt2$, since the differential has residues $\pm i\sqrt2$ at the two lifts of $\xi=0$. With these conventions and those of Section~\ref{sec:normal-radial} on the normal side, the leading classical comparison can be established analytically,
\begin{equation}
 \Pi_{B,*}^{(0)}=i\sqrt2\,\Pi_0(1^-).
 \label{eq:HS-leading-bridge}
\end{equation}
The original extremal limiting quartic is birationally isomorphic to the normal limiting cubic $Y^2=u(u^2-5u+8)$. Under this map the extremal $B$ cycle becomes the compact normal cycle $\gamma_A$. The normalized classical differentials differ by a single valued meromorphic exact form and a logarithmic differential. The exact term has zero period, while the logarithmic term has period $4\pi i$, precisely reproducing the horizon contribution in~\eqref{eq:outer-residue}. Appendix~\ref{app:HS-bridge} gives the map, the differential identity and the winding argument.

Numerically $\Pi_{B,*}^{(0)}=-17.3347925\ldots$, which gives the leading frequency coefficient $-0.512597186i$. To obtain the higher terms, we expand each $\Pi_B^{(k)}$ about the limiting point, insert the normal angular series into~\eqref{eq:HS-dictionary}, and invert~\eqref{eq:HS-EQC} in inverse powers of $n_{\rm ext}+1/2$. For $(\ell,m)=(2,-2)$, we obtain
\begin{multline}
 \Omega=-0.512597186i\left(n_{\rm ext}+\frac12\right)+(0.206813761+0.154683304i)\\
 +\frac{0.747383464+0.682481767i}{n_{\rm ext}+1/2}+\frac{2.83848011+0.108568953i}{(n_{\rm ext}+1/2)^2}\\
 +\frac{7.66189950+3.25261236i}{(n_{\rm ext}+1/2)^3}+\frac{17.9604349+15.4619794i}{(n_{\rm ext}+1/2)^4}+O((n_{\rm ext}+1/2)^{-5}).
 \label{eq:HS-example}
\end{multline}
The real constant term $0.206813761$ corresponds to $\operatorname{Re}\omega\to-m\times0.0517034404/M$. This agrees with the extremal limit $\hat\omega(a\to M)\simeq0.051704/M$ of the leading real frequency reported in~\cite{KH}, where $\omega\simeq-m\hat\omega$ at leading order.

For this example, we have also checked, at the first several orders in $w=\Omega^{-1}$, the agreement at the level of the quantization conditions:
\begin{equation}
 \frac{w\Pi_B}{\hbar_{\rm ext}}=\sum_{j\ge0}\Pi_j(1^-)\,w^j,
 \label{eq:HS-period-level-check}
\end{equation}
with the same branch conventions, order by order in $w$ up to the orders compared.

The extremal and non-extremal scalar frequency coefficients, written in terms of $n_{\rm ext}$ and $N$ respectively, agree at all the orders we have compared in the tested sectors $(\ell,m)=(2,-2),(2,-1),(2,0),(3,-1)$. Since the quantum periods in~\cite{HS} are known to high orders, this comparison can be continued further. For $\bar L>0$, the scalar $b_2$ in~\eqref{eq:scalar-b2} is needed. We have also checked that $n_{\rm ext}=N=n$ at $n=10,20,20$ for the three nonaxisymmetric sectors, respectively. In the axisymmetric $(2,0)$ sector, $n_{\rm ext}=n$ was checked, but $N$ was not fixed independently.\footnote{We exclude the case $m=\ell>0$, for which the Stokes geometry of the extremal problem is different~\cite{HS}.} We stress that the agreement above concerns the coefficients. We do not claim that the two boundary value problems or their Stokes graphs are equivalent, and we do not discuss the interchange of the limits $N\to\infty$ and $\chi\to1$ in general.
\section{Conclusions}
\label{sec:conclusion}
In this paper, we developed a systematic method to compute the high overtone expansion of Kerr QNM frequencies. We first computed the large-$c$ expansion of the angular eigenvalue by a simple polynomial recursion. Inserting it into the dual radial equation, we found that the WKB expansion is defined on a fixed genus-one spectral curve, and that the quantum periods at the second and higher orders reduce to two elliptic periods. This allows us to determine the frequency coefficients to any finite order. We tested the truncated expansions against QNM frequencies obtained by solving the angular and radial equations directly, and found good agreement for all three types of perturbations. For $\chi=4/5$ and $N=120$, the errors reach $10^{-21}$. We also found that, in the tested scalar cases, the coefficients agree with those obtained from the exact quantization condition for extremal Kerr black holes~\cite{HS}. At the classical level, this agreement was proved analytically, and the horizon contribution $4\pi i$ appears naturally from a logarithmic term.

Our analysis also revealed some limitations of the fixed spin expansion. In the small spin limit, $d_1$ and $d_2$ diverge as fractional powers of $\chi$, and the expansion is reliable only for $N\gg\chi^{-4/3}$ in the gravitational example. Moreover, the coefficients grow factorially, and the series is asymptotic.

Let us mention some directions for future work.
Since our series is asymptotic, an important problem is its resurgent structure. The least-term order $J_{\rm LT}$ grows roughly in proportion to $N$, which suggests that it is controlled by the Borel singularities closest to the origin. Locating these singularities would also clarify the corresponding exponentially small corrections and the relation between $J_{\rm LT}$ and $J_{\rm best}$. For Schwarzschild and Reissner--Nordstr\"om black holes, the WKB series at fixed overtone number was found to be Borel summable~\cite{HatsudaBorel}. The inverse overtone expansion considered here is a different series, and its Borel summability requires a separate analysis.

The small spin regime also deserves further study. A uniform treatment for $N\lesssim\chi^{-4/3}$ would connect our expansion with the Schwarzschild asymptotics and the transition discussed in~\cite{DGM}. For Schwarzschild black holes themselves, the corrections to the leading asymptotic form have been computed only to low orders~\cite{Maassen,MusiriSiopsis,CardosoLemosYoshida,KaoSchwarzschild}, and a systematic construction of the high overtone expansion is still lacking.

Near extremality, the branch point $u_0$ approaches a puncture. Although our coefficients agree with those from the extremal quantization condition~\cite{HS}, a uniform description would clarify how the two limits are related. For this purpose, it is desirable to obtain an exact quantization condition for non-extremal Kerr black holes by exact WKB analysis, as was done in the extremal case~\cite{HS}. Such an analysis would also put the all-order quantization condition~\eqref{eq:normal-quantization}, which we assumed in this paper, on a firm basis.

We have restricted ourselves to the normal angular branches. The anomalous branches~\cite{KBM,VC} require a separate analysis, including the question of which branch the physical modes follow at fixed spin as the overtone number increases. The WKB integer $N$ and the overtone number $n$ can differ, because low-lying modes join or leave the asymptotic sequence through the branch cut. A general understanding of this relation is also lacking.

In the gauge theory description, an expansion around a strong coupling point might improve the convergence at high overtones, as mentioned in the Introduction. It is an interesting question whether our coefficients can be reproduced from such an expansion of the free energy in the Nekrasov--Shatashvili limit~\cite{AGH}.

High overtone spectra can also be very sensitive to deformations of the black hole potential~\cite{MNNOInstability}. The present method may be useful to examine such effects for Kerr black holes.
\section*{Data availability}
A Mathematica package that evaluates the expansion~\eqref{eq:intro-series} for the normal branches, together with the exact coefficients of the periods $\Pi_j$ up to $j=25$ and of the angular eigenvalue up to $\mathcal A^{\rm nor}_{39}$, is available at Zenodo, \href{https://doi.org/10.5281/zenodo.23691045}{doi:10.5281/zenodo.23691045}. The other numerical data used in this paper are available from the author upon reasonable request.

\section*{Acknowledgments}
I used ChatGPT (GPT-5.6 and GPT-6, OpenAI) and Claude (Claude Opus 5.5 and Claude Fable 5.1, Anthropic) for language editing and assistance with analytic and numerical calculations, and Codex (OpenAI) and Claude Code (Anthropic) for code generation. I checked all the outputs. This work was partially supported by JSPS KAKENHI Grant Nos.~23K25790 and 26K07110.
\appendix
\section{Analytic constructions}
\label{app:analytic}
In this appendix, we give three analytic constructions used in the main text: the algebraic origin of the dual radial equation, the polynomial implementation of the quantum periods and the classical relation to the extremal period.

\subsection{Dual radial equation and boundary adapted series}
\label{app:dual}
To compare the two radial formulations, start with the homogeneous Teukolsky equation~\cite{Teukolsky,Leaver},
\begin{align}
 &\Delta^{-s}(\Delta^{s+1}R')'
 +\left[\frac{K_r^2-2i s(r-M)K_r}{\Delta}
       +4i s\omega r-A-c^2+2mc\right]R=0,\nonumber\\
 &K_r=(r^2+a^2)\omega-am.
 \label{eq:standard radial-app}
\end{align}
Here $r$ is the Boyer--Lindquist radius, $R(r)$ is the standard radial wave function, $\Delta=(r-r_+)(r-r_-)$, and $r_+=M(1+\rho)$ and $r_-=M(1-\rho)$ are the radii of the outer and inner horizons. A prime in this equation denotes $r$ differentiation. We introduce the coordinate
\begin{equation}
 z_T=\frac{r-r_-}{2M\rho},
 \label{eq:standard-CHE-coordinate}
\end{equation}
and remove the first derivative by $\psi_T=\Delta^{(s+1)/2}R$. Then we obtain the symmetric confluent Heun form~\cite{CasalsTdC}
\begin{equation}
\begin{aligned}
0={}&\Biggl[z_T(z_T-1)\partial_{z_T}^2-p^2z_T(z_T-1)
       -\mu_3p(2z_T-1)+E+\frac14\\
 &\hspace{2.2em}-\frac{\mu_1\mu_2}{z_T}
       -\frac{(\mu_1+\mu_2)^2-1}{4z_T(z_T-1)}\Biggr]\psi_T ,
\end{aligned}
 \label{eq:symmetric-CHE}
\end{equation}
with
\begin{align}
 \mu_1&=s+i\Omega,&
 \mu_2&=\frac{i(\Omega-\chi m)}{\rho},&
 \mu_3&=-s+i\Omega,\nonumber\\
 p&=i\rho\Omega,&
 E&=-A-s(s+1)+2\Omega^2-c^2-\frac14.
 \label{eq:CHE-dictionary}
\end{align}
Here $\mu_i$ are the Heun parameters, $p$ fixes the irregular exponential scale, and $E$ is the accessory parameter. In the Schwarzschild limit, the same exchange $\mu_2\leftrightarrow\mu_3$ was noted in~\cite{AGH} to relate the Teukolsky and Regge--Wheeler dictionaries. For Kerr, we exchange $\mu_2$ and $\mu_3$ in~\eqref{eq:symmetric-CHE}, relabel the independent variable as $z$, and set $\psi_{\rm d}=\sqrt{f}\,\phi$. Here $\psi_{\rm d}$ is the normal form solution of the exchanged equation, and $\phi$ is the dual wave function in Section~\ref{sec:dual-radial}. The resulting differential operator is exactly~\eqref{eq:alt}. Note that this is an identity of differential operators after the mass exchange, and it does not identify $z$ with $z_T$. Indeed, division of the exchanged equation by $z(z-1)$ gives its normal form coefficient, whereas the dual equation gives the coefficient $Q_{\rm dual}^{\rm NF}$ in normal form (NF),
\[
 Q_{\rm dual}^{\rm NF}=\frac{\Omega^2-V_{\rm d}}{f^2}
                   -\frac{f''}{2f}+\frac{(f')^2}{4f^2}.
\]
Primes in this display denote $z$ derivatives. The two rational functions agree after $c=\chi\Omega/2$ and $\rho^2=1-\chi^2$ are used.

The associated boundary factors can also be compared explicitly. For an ordering $(a,b,d)$ of the three Heun parameters, the Jaff\'e series is
\begin{equation}
 \psi_{\rm d}=e^{p(z-1)}z^{(a+b+2d-1)/2}(z-1)^{-(a+b-1)/2}
              \sum_{j\ge0}b_j\left(\frac{z-1}{z}\right)^j.
 \label{eq:Jaffe-generic}
\end{equation}
Here $b_j$ are Jaff\'e series coefficients. With $(a,b,d)=(\mu_1,\mu_3,\mu_2)$, division by $\sqrt f$ yields
\begin{equation}
 \phi=e^{p(z-1)}z^{i\Omega+\mu_2}(z-1)^{-i\Omega}
                     \sum_{j\ge0}b_j\zeta_{\rm r}^j,
 \qquad \zeta_{\rm r}=1-z^{-1}.
 \label{eq:dual-Jaffe}
\end{equation}
The horizon factor is $(z-1)^{-i\Omega}$. At infinity the explicit prefactor is $e^{pz}z^{\mu_2}$, up to a constant, as in~\eqref{eq:inf}. Therefore the power multiplying the exponential is fixed by the same dictionary as the differential equation.

These exponents combine the data of both Kerr horizons. With the surface gravities $\kappa_\pm=(r_+-r_-)/[2(r_\pm^2+a^2)]$ and angular velocities $\Omega^{\rm H}_\pm=a/(r_\pm^2+a^2)$ of the outer and inner horizons, one has $\kappa_+^{-1}-\kappa_-^{-1}=4M$ and $\Omega^{\rm H}_+/\kappa_+=\Omega^{\rm H}_-/\kappa_-$. Here both $\kappa_+$ and $\kappa_-$ are taken to be positive. Hence
\begin{equation}
 \Omega=\frac12\left[\frac{\omega-m\Omega^{\rm H}_+}{\kappa_+}-\frac{\omega-m\Omega^{\rm H}_-}{\kappa_-}\right],\qquad
 \frac{\Omega-\chi m}{\rho}=\frac12\left[\frac{\omega-m\Omega^{\rm H}_+}{\kappa_+}+\frac{\omega-m\Omega^{\rm H}_-}{\kappa_-}\right].
 \label{eq:dual-exponents-horizons}
\end{equation}
The horizon exponent in~\eqref{eq:dual-Jaffe} therefore corresponds to the difference of the outer and inner horizon data, with effective surface gravity $1/(4M)$ and no rotation, whereas the Coulomb-type exponent at infinity in~\eqref{eq:inf} corresponds to their sum. Analogous combinations of the horizon data determine the surface gravities and angular velocities of the dual Kerr--de Sitter spacetime in~\cite{HIZ}.

For completeness, let $\sigma_k^{\mu}$ denote the elementary symmetric polynomials of $\mu_1,\mu_2,\mu_3$. In the characteristic continued fraction, the diagonal coefficient and adjacent product are
\begin{align}
 A_j^{(0)}&=E+\frac14+2j(p-j)
             +(2j+1-p)(\sigma_1^\mu-1)-\sigma_2^\mu,
             \label{eq:symmetric-diagonal}\\
 A_{j-1}^{(+)}A_j^{(-)}
   &=j\{(j-\sigma_1^\mu)[j(j-\sigma_1^\mu)+\sigma_2^\mu]
                          +\sigma_3^\mu\}.\nonumber
\end{align}
Their permutation symmetry explains why the boundary adapted characteristic recurrence survives the exchange. The subtraction of $1$ in~\eqref{eq:symmetric-diagonal} is required by direct substitution. The literal coefficients used for numerical evaluation are given in Appendix~\ref{app:direct}.

Let us comment on the spectral equivalence. Whiting's integral transformation and its extensions identify the exchanged problem in the mode stability domain. The point spectrum result in~\cite{CasalsTdC} uses $\operatorname{Re}p\le0$, which corresponds to $\operatorname{Im}\omega\ge0$~\cite{Whiting,TdC}. For QNM frequencies, the continuation of the characteristic function of the minimal solution, together with the comparison of the two radial formulations, strongly supports the isospectrality.

\subsection{Polynomial pair recursion and a worked period reduction}
\label{app:polynomial}
The direct recurrence~\eqref{eq:Riccati-recursion} is sufficient at low order, but repeated differentiation of algebraic fractions becomes costly. The following representation keeps only polynomial operations while retaining the unreduced momentum needed by the next order. Write $G=G_\sigma$ and, for $j\ge1$, define
\begin{equation}
 a_j=\left\lfloor\frac{3j-1}{2}\right\rfloor,
 \qquad b_j=\left\lfloor\frac{3j}{2}\right\rfloor,
 \qquad P_j=\frac{e_j(u)}{G^{a_j}}+\frac{o_j(u)Y}{G^{b_j}}.
 \label{eq:pair-definition}
\end{equation}
The symbols $a_j,b_j$ in this subsection are denominator exponents, distinct from the scalar angular coefficients and the Jaff\'e coefficients used above. At $j=0$ set $a_0=b_0=0$, $e_0=0$, $o_0=1/2$. The polynomial $e_j$ represents the branch even part and $o_j$ the branch odd part. Introduce the two derivative operations
\begin{equation}
 E_a[e]=F(Ge'-aG'e),\qquad
 O_b[o]=F\left[Go'+\left(\frac12-b\right)G'o\right].
 \label{eq:pair-derivatives}
\end{equation}
They satisfy $D(e/G^a)=E_a[e]/G^{a+1}$ and $D(oY/G^b)=O_b[o]Y/G^{b+1}$. Splitting~\eqref{eq:Riccati-recursion} into its branch parities therefore gives
\begin{align}
 e_j={}&i G^{a_j-b_{j-1}-1}O_{b_{j-1}}[o_{j-1}]
  -\sum_{r=1}^{j-1}\left[
    G^{a_j-a_r-b_{j-r}}e_ro_{j-r}
   +G^{a_j-b_r-a_{j-r}}o_re_{j-r}\right],\nonumber\\
 o_j={}&G^{b_j-1}Q_j+i G^{b_j-a_{j-1}-2}E_{a_{j-1}}[e_{j-1}]
  -\sum_{r=1}^{j-1}\left[
    G^{b_j-1-a_r-a_{j-r}}e_re_{j-r}
   +G^{b_j-b_r-b_{j-r}}o_ro_{j-r}\right].
 \label{eq:pair-recursion}
\end{align}
Terms multiplying an identically zero polynomial are omitted before powers are formed. In particular, the derivative of $e_0=0$ is absent. All remaining powers in this implementation are nonnegative. The vanishing factors of~\eqref{eq:radial-Qj} give $F\mid e_j$ for $j\ge1$ and $F\mid o_j$ for $j\ge2$. Hence the raw differential is
\begin{equation}
 \lambda_j=R_j(u)\frac{du}{Y^{2k_j+1}},\qquad
 R_j=\frac{o_j}{F},\quad k_j=b_j-1,\qquad j\ge2.
 \label{eq:raw-pair-output}
\end{equation}
Its period is then evaluated by the moment relations in Section~\ref{sec:period-reduction}. A representative modulo exact differentials can be used for the integration, but not in the Riccati recursion.

Let us illustrate this with a concrete example. The second order numerator obtained directly from~\eqref{eq:P2} is
\begin{equation}
 R_2=\frac{Q_2}{F}G^2-\frac{F'G'+FG''}{4}G
                     +\frac{5F(G')^2}{16}-\frac{Q_1^2}{F}G.
 \label{eq:R2-worked}
\end{equation}
Although written as a rational expression, it is polynomial because $Q_2$ is divisible by $F$ and $Q_1^2/F$ is polynomial. To reduce $R_2du/Y^5$, one may either use the moments or find a polynomial $X$ and two constants $H_0,H_1$ satisfying
\begin{equation}
 R_2-(H_0+H_1u)G^2=GX'-\frac32G'X.
 \label{eq:reduction-certificate}
\end{equation}
The right hand side is the numerator of $d(X/Y^3)$. Equating powers of $u$ gives a finite linear system. A polynomial $X$ of degree five suffices. For the scalar example $(s,m,\bar L)=(0,-2,0)$, $Q_2=F(13/4+u)$ and the answer is
\begin{align}
 H_0&=\frac{4\sigma^2+(170-96i)\sigma+522-864i}{24(\sigma+27)},\nonumber\\
 H_1&=-\frac{2\sigma^2+(82-48i)\sigma-108}{24(\sigma+27)}.
 \label{eq:Pi2-worked}
\end{align}
Integrating~\eqref{eq:reduction-certificate} around $\gamma_A$ removes the exact term, so $\Pi_2^{\gamma_A}=H_0J_0+H_1J_1$ and $d_1=-(H_0J_0+H_1J_1)/(2\pi)$. The same reduction applies at every finite order, and no new integration basis is needed.

\subsection{Classical period bridge at the extremal endpoint}
\label{app:HS-bridge}
In this subsection, we prove the leading identity used in Section~\ref{sec:HS}. At $\chi=1$ the normal curve and differential are
\begin{equation}
 E_N:\quad Y^2=u(u^2-5u+8),\qquad
 \lambda_0=\frac{Y}{2u^2(u-1)}\,du.
 \label{eq:bridge-normal-curve}
\end{equation}
Let $\xi$ denote the rescaled extremal coordinate in~\cite{HS}. At $(\mathcal A,\mu)=(1/\sqrt2,\sqrt2)$ the original extremal classical curve and differential can be written
\begin{align}
 E_H:\quad R_H^2&=(\xi+\sqrt2)(\xi+1/\sqrt2)
      (\xi^2+\xi/\sqrt2+1),\nonumber\\
 \eta_H&=i\frac{R_H}{\xi^2}\,d\xi.
 \label{eq:bridge-HS-curve}
\end{align}
The direct birational map and its inverse are
\begin{align}
 u&=\frac{2(\xi+\sqrt2)}{\xi+1/\sqrt2},&
 Y&=\frac{2R_H}{(\xi+1/\sqrt2)^2},\nonumber\\
 \xi&=\frac{4-u}{\sqrt2(u-2)},&
 R_H&=\frac{Y}{(u-2)^2}.
 \label{eq:bridge-birational-map}
\end{align}
Indeed $G(u(\xi))=4R_H^2/(\xi+1/\sqrt2)^4$, so the smooth compact curves are isomorphic. Both have modular invariant $j=-4/7$. The extremal turning points map to $u_+$ and $u_-$, and the $B$ cycle $\gamma_{\beta_-\beta_+}$ maps to the compact normal cycle $\gamma_A$ with the orientation of Section~\ref{sec:radial-geometry}.

We normalize and pull back the extremal differential as
\begin{equation}
 \widetilde\lambda_H:=\frac{\eta_H}{i\sqrt2}
 =-\frac{2Y}{(u-2)^2(u-4)^2}\,du.
 \label{eq:bridge-pulled-differential}
\end{equation}
Define the meromorphic functions
\begin{equation}
 \mathcal R=\frac{(3u-8)Y}{u(u-2)(u-4)},\qquad
 \mathcal G=\frac{Y+u(3-u)}{Y-u(3-u)}.
 \label{eq:bridge-RG}
\end{equation}
Using $dY=G'(u)du/(2Y)$ and $G-u^2(3-u)^2=-u(u-1)(u-2)(u-4)$ gives the exact differential identity
\begin{equation}
 \widetilde\lambda_H=\lambda_0+d\mathcal R+d\log\mathcal G.
 \label{eq:bridge-differential-certificate}
\end{equation}
The function $\mathcal R$ is single valued, hence its closed period vanishes. The logarithm is not globally single valued. On the chosen lift from $u_+$ to $u_-$, the image of $\mathcal G$ winds once counterclockwise around the origin. The reversed involution image contributes the same winding because $\iota^*\mathcal G=1/\mathcal G$. Therefore
\begin{equation}
 \oint_{\gamma_A}d\log\mathcal G=4\pi i.
 \label{eq:bridge-log-period}
\end{equation}
Equivalently, one can parametrize the outgoing leg by $u=5/2+i t$, $\sqrt7/2\ge t\ge-\sqrt7/2$: $\mathcal G$ runs from $-1$ through the lower half plane, crosses the positive real axis and returns to $-1$ through the upper half plane, so its argument increases by $2\pi$ on that leg and by another $2\pi$ on the second sheet.

Integrating~\eqref{eq:bridge-differential-certificate} yields
\begin{equation}
 \frac{\Pi_{B,*}^{(0)}}{i\sqrt2}
 =\Pi_0^{\gamma_A}(1)+4\pi i
 =\Pi_0(1),
 \label{eq:bridge-integrated}
\end{equation}
which is~\eqref{eq:HS-leading-bridge}. The compact period has a regular one-sided limit because the transported roots remain separated and the corresponding integrand on the compact cycle is uniformly nonsingular for $\chi$ sufficiently close to one. Therefore the endpoint value in~\eqref{eq:bridge-integrated} equals $\Pi_0(1^-)$.

Note that the argument above proves only the leading classical identity. The agreement at higher orders,~\eqref{eq:HS-period-level-check}, was checked for $(\ell,m)=(2,-2)$.
\section{Numerical methods and validation}
\label{app:numerical}
In this appendix, we summarize the numerical methods used in Section~\ref{sec:normal-checks}. The quoted uncertainties are estimated by varying the numerical parameters, and are not rigorous error bounds.

\subsection{Direct computation of QNM frequencies}
\label{app:direct}
For each trial frequency $\Omega$, we solve the angular problem at $c=\chi\Omega/2$ by the spectral method in the basis of spin-weighted spherical harmonics~\cite{CZ}. For the gravitational modes, we also use Leaver's continued fraction for the angular problem~\cite{Leaver}. The eigenvalue is transported continuously from $c=0$, and the resulting value, not its large-$c$ series, is used in the radial problem. For the dual radial equation, the series~\eqref{eq:dual-Jaffe} gives the three-term recurrence $\alpha_jb_{j+1}+\beta_jb_j+\gamma_jb_{j-1}=0$ with $b_{-1}=0$, where
\begin{align}
 \alpha_j={}&\rho^2(j+1)(j+1-2i\Omega),\nonumber\\
 \beta_j={}&\rho[(1+\rho)\Omega+i j]
       [(1+\rho)(2+\rho)\Omega-2\chi m+i\rho(j+2)]\nonumber\\
 &-i\chi\rho[m-\chi\Omega+i\rho m\Omega]\nonumber\\
 &-\rho^2[j^2+A+s+1+(1+\rho)^2\Omega^2+\chi^2\Omega^2/4],\nonumber\\
 \gamma_j={}&-[(1+\rho)\Omega-\chi m+i\rho j]^2-\rho^2s^2.
 \label{eq:literal-CF}
\end{align}
The same recurrence was obtained in~\cite{HatsudaAlt} in a different notation. The QNM condition requires $b_j$ to be the minimal solution, which leads to Leaver's continued fraction equation~\cite{Leaver}. To compute high overtones accurately, we use Leaver's inversion of the continued fraction and Nollert's asymptotic tail~\cite{Nollert1993}. We also solve the recurrence for the standard radial Teukolsky equation as a check. For the electromagnetic modes, for instance, the two radial formulations agree within $1.66\times10^{-55}$ at $(\chi,N)=(1/2,100)$ and $1.09\times10^{-60}$ at $(4/5,120)$.

Representative final settings are listed in Table~\ref{tab:numerical-settings}. For each root, we vary the working precision, the angular cutoff, and the depth and tail order of the continued fraction, and estimate the numerical uncertainty $u_\Omega$. As shown in Table~\ref{tab:root-uncertainty}, $u_\Omega$ is much smaller than the minimal truncation error in all the cases.
\begin{table}[tbp]
 \centering
 \caption{Representative final numerical settings for the direct roots in Figure~\ref{fig:three-spins}.}
 \label{tab:numerical-settings}
 \begin{tabular}{lccc}\toprule
 & $s=0$ & $s=-1$ & $s=-2$\\\midrule
 working precision (digits) & 140 & 110 & 100 or 125\\
 radial continued fraction depth & 30000 & 20000 & 14000 or 22000\\
 tail order & 38 & 32 & 24 or 30\\
 angular basis cutoff & 120 & 100 & 70 or 95\\
 angular continued fraction depth & --- & --- & 400 or 560\\\bottomrule
 \end{tabular}
\end{table}
\begin{table}[tbp]
 \centering
 \caption{Numerical uncertainties $u_\Omega$ of representative direct roots with $(\ell,m,\bar L)=(2,-2,0)$, compared with the minimal truncation errors $E_{J_{\rm best}}$ in Table~\ref{tab:normal-truncation}.}
 \label{tab:root-uncertainty}
 \begin{tabular}{rrrrr}\toprule
 $s$&$\chi$&$N$&$u_\Omega$&$E_{J_{\rm best}}$\\\midrule
 $0$&$4/5$&120&$7.27\times10^{-50}$&$2.71\times10^{-21}$\\
 $-1$&$1/2$&40&$3.43\times10^{-43}$&$5.27\times10^{-4}$\\
 $-1$&$4/5$&120&$5.72\times10^{-48}$&$3.02\times10^{-21}$\\
 $-2$&$1/2$&40&$1.17\times10^{-30}$&$2.68\times10^{-2}$\\
 $-2$&$4/5$&120&$1.26\times10^{-49}$&$4.25\times10^{-21}$\\
 $-2$&$19/20$&120&$3.26\times10^{-41}$&$3.67\times10^{-25}$\\\bottomrule
 \end{tabular}
\end{table}

\subsection{Overtone numbers and coefficients}
\label{app:normal-numerics}
The integer $N$ labels the roots in Section~\ref{sec:normal-checks}, and it should be distinguished from the overtone number $n$. We define $n$ as in the standard tables~\cite{Leaver,BCS2009}. The overtones are numbered in the Schwarzschild limit, keeping the slot of the algebraically special frequency (for example, $n=8$ at $\Omega=-4i$ for $s=-2$ and $\ell=2$), and the numbers are carried along each sequence as $\chi$ increases~\cite{CZ,CZ2016}. The roots are zeros of the continued fraction with the principal branch of the Nollert tail, whose branch cut lies on the negative imaginary $\Omega$ axis. They are the poles of the Green function on the physical sheet.

The two labels can differ, because low-lying modes need not belong to the asymptotic sequence described by~\eqref{eq:intro-series}. Some modes reach the negative imaginary axis and pass through the branch cut onto the unphysical sheet for a finite range of $\chi$~\cite{CZ2016}, while some zeros on the unphysical sheet come onto the physical sheet. When such a mode leaves or joins the asymptotic sequence, $n-N$ changes, by one in the cases we have examined. Therefore, $n-N$ depends on the mode, and in some cases also on $N$. A count of the roots on the physical sheet at fixed $\chi$ can also differ from $n$.

For the modes in Section~\ref{sec:normal-checks}, we find $n=N$ for $s=0$ and $s=-1$, and $n=N+1$ for $s=-2$. For all the roots in Section~\ref{sec:normal-checks}, the overtone numbers were obtained by continuation from the Schwarzschild limit. For $s=-2$ at $\chi=1/2$, this was also confirmed by a count of the roots by the argument principle. For the extremal comparison in Section~\ref{sec:HS}, we have similarly checked that $n_{\rm ext}=N=n$ for $(\ell,m,n)=(2,-2,10),(2,-1,20),(3,-1,20)$, and $n_{\rm ext}=n$ for $(\ell,m)=(2,0)$.

The coefficients $d_j$ are computed at least up to $d_{159}$. We have checked the angular and radial recursions at several truncation orders, and the elliptic periods at several working precisions. The periods in Figure~\ref{fig:rotation} were computed at 339 values of $\chi$ with 90 digits and recomputed with 60 digits, with estimated uncertainties below $10^{-34}$. The reduction of the higher periods was also checked by direct contour integration.
\FloatBarrier
\bibliographystyle{amsmod}
\bibliography{kerr_refs}
\end{document}

%% file: two_sheet_cycle_twopanel.tikz
% Figure 1: (a) the connected outer QNM cycle gamma_out and (b) its
% decomposition gamma_A + delta_1^(+) - delta_1^(-), on the two sheets of the leading normal curve at chi = 1/2.
% Input inside the figure environment (scaled to the text width). Requires \usepackage{ifthen} and the TikZ
% libraries arrows.meta, bending, decorations.markings, calc, positioning (loaded in the preamble).
% Points are natural coordinates (Re u, Im u). The oblique projection is set by sheetaxes.
% ----------------------------- colours --------------------------------
\definecolor{cutblue}{RGB}{35,70,200}      % branch cuts and branch points
\definecolor{punctred}{RGB}{205,30,30}     % punctures u = 0, 1
\definecolor{cyclegreen}{RGB}{0,125,95}    % the cycle

% ----------------------------- switches -------------------------------
\newif\ifshowglue     \showgluetrue     % dotted lines joining the sheets at u_+-
\newif\ifshowformula  \showformulatrue  % homology relation under the figure
\newif\ifshowgamma    \showgammatrue    % label gamma_out next to the cycle

% ------------------------------ labels --------------------------------
\newcommand{\sheetOneLabel}{sheet $+$}
\newcommand{\sheetOneNote}{$Y(1)=+2$}
\newcommand{\sheetTwoLabel}{sheet $-$}
\newcommand{\sheetTwoNote}{$Y(1)=-2$}

% --------------------------- geometry ---------------------------------
% oblique projection: x = Re u, y = Im u drawn as a tilted plane
\tikzset{sheetaxes/.style={x={(0.80cm,0cm)}, y={(0.23cm,0.26cm)}}}
% sheet extent in natural coordinates
\def\ReMin{-2.5}  \def\ReMax{4.8}
\def\ImMin{-3.1}  \def\ImMax{3.1}
% vertical offset between the two sheets
\def\sheetgap{2.55cm}

% points (schematic positions close to chi = 1/2:
%   u_0 = -0.90,  u_+- = 2.95 +- 2.15 i)
\def\uzero{(-0.90,0)}
\def\uplus{(2.95,2.15)}
\def\uminus{(2.95,-2.15)}
\def\pzero{(0,0)}      % puncture u = 0
\def\pone{(1,0)}       % puncture u = 1

% the cycle on one sheet: from u_+ around the LEFT of u = 1 to u_-
% (must enclose u = 1 but not u = 0)
\def\cyclepath{\uplus .. controls (1.25,2.65) and (0.42,1.35) .. (0.42,0)
                      .. controls (0.42,-1.35) and (1.25,-2.65) .. \uminus}
\def\cyclepathrev{\uminus .. controls (1.25,-2.65) and (0.42,-1.35) .. (0.42,0)
                         .. controls (0.42,1.35) and (1.25,2.65) .. \uplus}

% ------------------------------ styles --------------------------------
\tikzset{
  sheet/.style={line width=0.6pt, line join=round, fill=black!2},
  cut/.style={cutblue, dashed, line width=0.9pt, dash pattern=on 3.2pt off 2.2pt},
  bpoint/.style={circle, fill=cutblue, inner sep=0pt, minimum size=4.2pt},
  cycle/.style={cyclegreen, line width=1.15pt},
  glue/.style={black!45, densely dotted, line width=0.6pt},
  arrowat/.style={postaction={decorate}, decoration={markings,
      mark=at position #1 with {\arrow{Stealth[length=2.6mm,width=2.0mm]}}}},
  lab/.style={font=\small},
}
% red cross for a puncture
\tikzset{uplabel/.style={right=5pt}, umlabel/.style={right=5pt}, punctlabel/.style={below=3pt}}
\newcommand{\puncture}[1]{%
  \draw[punctred, line width=0.9pt] #1 ++(-2.3pt,-2.3pt) -- ++(4.6pt,4.6pt)
                                    #1 ++(-2.3pt,2.3pt)  -- ++(4.6pt,-4.6pt);}

% one sheet: frame, u-plane symbol, cuts, points (cycle drawn separately)
\newcommand{\drawsheet}[2]{%
  \draw[sheet] (\ReMin,\ImMin) -- (\ReMax,\ImMin) -- (\ReMax,\ImMax)
               -- (\ReMin,\ImMax) -- cycle;
  % u-plane symbol in the upper-right corner: its two lines follow the axes of the tilted plane
  \draw[line width=0.45pt] (\ReMax-0.56,\ImMax-0.40) -- (\ReMax-0.56,\ImMax-1.10)
                           -- (\ReMax-0.12,\ImMax-1.10);
  \node[lab, inner sep=0pt] at (\ReMax-0.31,\ImMax-0.70) {$u$};
  % branch cuts
  \draw[cut] (\ReMin,0) -- \uzero;
  \draw[cut] \uminus -- \uplus;
  % branch points
  \node[bpoint] at \uzero {};  \node[lab, below right=0pt and -1pt] at \uzero {$u_0$};
  \node[bpoint] at \uplus {};  \node[lab, uplabel] at \uplus {$u_+$};
  \node[bpoint] at \uminus {}; \node[lab, umlabel] at \uminus {$u_-$};
  % punctures
  \puncture{\pzero} \node[lab, punctlabel] at \pzero {$0$};
  \puncture{\pone}  \node[lab, punctlabel] at \pone  {$1$};

}

% small loop around the puncture at (#1,0): a circle of radius \loopr in the u plane, drawn in the
% tilted sheet coordinates (so it appears as the oblique view of a circle). #2 = ccw or cw.
% The arrow is bent along the loop and placed on its front part, where the drawn curve is flattest.
\def\loopr{0.62}
\newcommand{\loopcircle}[2]{%
  \draw[cycle] (#1,0) circle[radius=\loopr];
  \ifthenelse{\equal{#2}{ccw}}%
    {\draw[cycle, -{Stealth[length=2.2mm,width=1.8mm,bend]}] (#1,0) ++(245:\loopr) arc[start angle=245, end angle=295, radius=\loopr];}%
    {\draw[cycle, -{Stealth[length=2.2mm,width=1.8mm,bend]}] (#1,0) ++(295:\loopr) arc[start angle=295, end angle=245, radius=\loopr];}%
}
\def\panelshift{7.9cm}
\begin{tikzpicture}
  % ---------- coordinates for gluing lines, both panels ----------
  \foreach \px/\tag in {0cm/a, \panelshift/b}{
    \begin{scope}[xshift=\px, sheetaxes]
      \path \uplus coordinate (Up1\tag) \uminus coordinate (Um1\tag);
    \end{scope}
    \begin{scope}[xshift=\px, yshift=-\sheetgap, sheetaxes]
      \path \uplus coordinate (Up2\tag) \uminus coordinate (Um2\tag);
    \end{scope}
  }
  % ================= panel (a): connected representative =================
  \begin{scope}[yshift=-\sheetgap, sheetaxes]
    \drawsheet{}{}
    \draw[cycle, arrowat=0.76] \cyclepathrev;
    \node[bpoint] at \uplus {}; \node[bpoint] at \uminus {};
    \path (1.05,1.77) coordinate (CycBota);   % a point on the cycle (upper arc)
  \end{scope}
  \draw[glue] (Up1a) -- (Up2a); \draw[glue] (Um1a) -- (Um2a);
  \begin{scope}[sheetaxes]
    \drawsheet{}{}
    \draw[cycle, arrowat=0.24] \cyclepath;
    \node[bpoint] at \uplus {}; \node[bpoint] at \uminus {};
    \node[lab, anchor=south west] at (\ReMin,\ImMax) {(a)};
    \path (2.171,-2.258) coordinate (CycTopa); % a point on the cycle (lower arc), chosen so that
                                               % the two leaders below are mirror images of each other
  \end{scope}
  % label of the cycle between the two sheets, with thin leaders to the parts on both sheets
  \node[lab, text=cyclegreen, inner sep=1.5pt] (gout) at ($(Um1a)!0.5!(Up2a)+(-1.3cm,0)$) {$\gamma_{\rm out}$};
  \draw[cyclegreen, line width=0.35pt, shorten >=1.2pt] (gout.north) -- (CycTopa);
  \draw[cyclegreen, line width=0.35pt, shorten >=1.2pt] (gout.south) -- (CycBota);
  % ================= panel (b): decomposition gamma_A + delta_1^(+) - delta_1^(-) =================
  \begin{scope}[xshift=\panelshift, yshift=-\sheetgap, sheetaxes]
    \tikzset{punctlabel/.style={below=7pt}}
    \drawsheet{}{}
    % -delta_1^(-): clockwise around u = 1 on sheet -
    \loopcircle{1}{cw}
    \node[lab, text=cyclegreen] at (1.0,1.75) {$-\delta_1^{(-)}$};
  \end{scope}
  \draw[glue] (Up1b) -- (Up2b); \draw[glue] (Um1b) -- (Um2b);
  \begin{scope}[xshift=\panelshift, sheetaxes]
    \tikzset{punctlabel/.style={below=7pt}, uplabel/.style={right=9pt}, umlabel/.style={right=10pt}}
    \drawsheet{}{}
    % gamma_A: counterclockwise around the cut [u_-, u_+] on sheet +
    % a capsule at constant distance \capd (on the page) from the cut, so that it touches neither
    % the cut nor the turning points; the arrow sits on its straight left side (counterclockwise)
    \def\capd{0.2cm}
    \draw[cycle] let \p1=(Um1b), \p2=(Up1b), \n1={atan2(\y2-\y1,\x2-\x1)} in
      ($(Um1b)+(\n1+90:\capd)$) -- ($(Up1b)+(\n1+90:\capd)$)
      arc[start angle=\n1+90, end angle=\n1-90, radius=\capd]
      -- ($(Um1b)+(\n1-90:\capd)$)
      arc[start angle=\n1-90, end angle=\n1-270, radius=\capd] -- cycle;
    \draw[cycle, arrowat=0.55] let \p1=(Um1b), \p2=(Up1b), \n1={atan2(\y2-\y1,\x2-\x1)} in
      ($(Up1b)+(\n1+90:\capd)$) -- ($(Um1b)+(\n1+90:\capd)$);
    \node[lab, text=cyclegreen] at (3.8,-0.4) {$\gamma_A$};
    % delta_1^(+): counterclockwise around u = 1 on sheet +
    \loopcircle{1}{ccw}
    \node[lab, text=cyclegreen] at (1.0,1.75) {$\delta_1^{(+)}$};
    \node[bpoint] at \uplus {}; \node[bpoint] at \uminus {};
    \node[lab, anchor=south west] at (\ReMin,\ImMax) {(b)};
    % sheet labels, once, at the far right
    \node[lab, anchor=west, align=left] at (\ReMax+0.75,-0.5) {\sheetOneLabel\\[-1pt]{\footnotesize \sheetOneNote}};
  \end{scope}
  \begin{scope}[xshift=\panelshift, yshift=-\sheetgap, sheetaxes]
    \node[lab, anchor=west, align=left] at (\ReMax+0.75,-0.5) {\sheetTwoLabel\\[-1pt]{\footnotesize \sheetTwoNote}};
    \path ({\ReMin},\ImMin) coordinate (BotB);
  \end{scope}
  \begin{scope}[yshift=-\sheetgap, sheetaxes]
    \path ({\ReMin},\ImMin) coordinate (BotA);
  \end{scope}
  % homology relation, centred under both panels
  \node[lab, anchor=north] at ($(BotA)!0.5!(BotB)+(3.6cm,-0.25cm)$)
      {$\gamma_{\rm out}\sim\gamma_A+\delta_1^{(+)}-\delta_1^{(-)}$};
\end{tikzpicture}